\documentclass{aa}  
\usepackage{graphicx}
\usepackage{txfonts}
\usepackage{hyperref}
\usepackage{diagbox}
\usepackage{makecell}
\usepackage{soul}

\usepackage{natbib}
\defcitealias{EHT1}{EHT L1}
\defcitealias{EHT3}{EHT L3}
\defcitealias{EHT4}{EHT L4}
\defcitealias{EHT5}{EHT L5}
\defcitealias{EHT6}{EHT L6}
\defcitealias{porth19}{Porth et al. 2019}

\usepackage{url}
\usepackage{color}
\usepackage{xcolor}
\usepackage{xspace}
\usepackage[normalem]{ulem}

\newcommand{\gy}{\textsc{Gyoto}\xspace}

\newcommand{\be}{\begin{equation}}
\newcommand{\ee}{\end{equation}}
\newcommand{\nn}{\nonumber}
\newcommand{\pp}{\varphi} 
\newcommand{\dd}{\mathrm{d}}

\graphicspath{{Figure/}}

\begin{document}

\title{Polarized signatures of orbiting hot spots: \\special relativity impact and probe of spacetime curvature}

\author{ 
  F. H. Vincent\inst{1}
  \and
  M. Wielgus\inst{2,3}
  \and
  N. Aimar\inst{1}
  \and
  T. Paumard\inst{1}
  \and
  G. Perrin\inst{1}
}

\institute{
  LESIA, Observatoire de Paris, Universit\'e PSL, CNRS, Sorbonne Universit\'es, UPMC Univ. Paris 06, Univ. de Paris, Sorbonne Paris Cit\'e, 5 place Jules Janssen, 92195 Meudon, France
  \email{frederic.vincent@obspm.fr}
  \and
  Max-Planck-Institut für Radioastronomie, Auf dem Hügel 69, 53121 Bonn, Germany   \email{maciek.wielgus@gmail.com}
  \and
Research Centre for Computational Physics and Data Processing, Institute of Physics, Silesian University in Opava, Bezru\v{c}ovo n\'am.~13, CZ-746\,01 Opava, Czech Republic
}

% \abstract{}{}{}{}{} 
% 5 {} token are mandatory

\abstract
{The Galactic Center supermassive black hole is well known to exhibit transient peaks of flux density on a daily basis across the spectrum.
 Recent infrared and millimeter observations have strengthened
 the case for the association between these flares and circular orbital motion
  in the vicinity of the event horizon. 
  The strongly polarized synchrotron radiation associated with these events leads to specific observables called QU loops, that is, looping motion in the Stokes QU plane of linear polarization.}
{We want to deepen the understanding of the QU loops associated with
  orbiting hot spots. We compute such loops in Minkowski and Schwarzschild spacetimes in order to determine which aspects of the observed patterns are due to special- or general-relativistic phenomena.
   }
  {We consider a parcel of energized plasma in circular motion in Minkowski spacetime, and in Keplerian orbit in the Schwarzschild spacetime. 
   We compute the polarized radiative transfer
    associated with this orbiting hot spot and derive the evolution of the flux density, astrometry, and Stokes Q and U parameters.
  }
  {We show that QU loops {in} Minkowski spacetime at low or
    moderate inclination $i \lesssim 45^\circ$ share all {qualitative}
    features of Schwarzschild QU loops: there exist QU loops for all
    setups considered (including face-on view and vertical
    magnetic field), there may be one or two QU loops per orbital period for a vertical magnetic field configuration, there are always two QU loops in case of a toroidal magnetic field. We provide analytical formulas in Minkowski spacetime to explain the details of this behavior. Moreover, we analyze the flux variation
    of the hot spot and show that it is dictated either by the angular dependence of the radiative transfer coefficients, or by relativistic
    beaming. In the former case, this can lead to extreme flux ratios
    even at moderate inclination. Finally, we highlight the increasing mirror asymmetry of the Schwarzschild QU track with increasing inclination and show that this behavior is a specific Schwarzschild feature caused by light bending. %\mw{maybe add "increasing MIRROR asymmetry"? Because asymmetry grows with increasing inclination in a sense that loop becomes less circular - deviation from a POLAR symmetry without relation to GR}
    %, the former effect being able to lead in the case of a
    %vertical field to a strong (factor $\approx 4$) flux modulation
    %at low inclination ($\approx 10^\circ)$.
   }
   {Although special-relativistic effects have not been extensively discussed in this context, they are a crucial part in generating the observed QU loops. However, general-relativistic light bending leads to a specific observable feature encoded in the asymmetry of the observed loops. This might allow quantifying the spacetime curvature.}


\maketitle
%
%-------------------------------------------------------------------

\section{Introduction}

The emission from the close surroundings of the
Galactic supermassive black hole Sagittarius~A* (Sgr~A*) is variable at all wavelengths, with a degree of variability that depends strongly on frequency~\citep{genzel10}. The source
exhibits local maxima of variable emission, from radio frequencies to X rays, called radiation flares~\citep[see e.g.][]{genzel10,morris23}.
The physical nature of these events remains unclear after
20 years of study since the first detected events~\citep{baganoff01,genzel03}.
Many models have been proposed, and we refer to~\citet{vincent14}
for a review. Among them, the class of
hot-spot models~\citep[][and references therein]{broderick06,hamaus09}
is of particular interest. The underlying assumption of this model is that
Sgr~A* flares are caused by the radiation emitted by transient,
localized (at least initially), compact (few gravitational radii),
orbiting (in the disk plane or along the jet funnel)
parcels of energized plasma in the inner region of the accretion/ejection flow
surrounding the black hole. This model is of particular relevance given the detections of orbital motions consistent with circular trajectories, very close to the event horizon, associated with infrared and X-ray flares~\citep{gravity18,wielgus22,gravity23}.
Such hot spots might be the end product of the acceleration of
particles in the inner regions of the flow by
magnetic reconnection~\citep[see e.g.][]{ripperda22,elmellah23}.
It has recently been shown that hot spots generated by magnetic
reconnection may account for photometric and astrometric infrared observations \citep{aimar23a}.

The polarization properties of infrared and millimeter flares have been studied since the early 2000s. \citet{eckart06} observed swings of the electric vector position angle (EVPA) of up to $40^\circ$ in 10~min during an infrared flare observed
by the NAOS/CONICA adaptive optics instrument, while~\citet{trippe07} measured a swing reaching $70^\circ$ within 15~min,
with the same instrument. The authors note that
these swings are consistent with a hot-spot model with an
orbital radius of the order of the innermost stable
circular orbit (ISCO) associated with the black hole, which corresponds to a Keplerian period of the order of 30~min for a non-spinning black hole of $\sim 4 \times 10^6\,M_{\odot}$. The change of polarization angle has been linked to the variation of the relative orientation between the direction of emission reaching the distant observer and that
of the ambient magnetic field, as the spot orbits around the black hole. The hot-spot model was further discussed in the context of these infrared polarized flare observations by~\citet{meyer06}.
Compatible infrared flare observations
and similar conclusions were obtained by~\citet{nishiyama09}.
At radio frequencies, \citet{marrone06} reported a $50^\circ$ EVPA swing over 2.5~hours during a millimeter flare observed by the Submillimeter Array, and noticed
a roughly periodic evolution of the angle with time.
When representing the evolution in the QU plane corresponding to the Stokes Q and U linear polarization
parameters, the authors obtained a loop pattern 
%\mw{[we just said it was roughly periodic in the last sentence] \st{associated
%with this roughly periodic evolution of the EVPA,}}
exhibiting two full orbits in the QU plane -- the first so-called QU loop reported in the literature. The authors argued that this signature might be associated with a hot spot orbiting at a radius larger than the ISCO of the black hole. Two different instruments observed QU loops recently:
the Very Large Telescope Interferometer GRAVITY beam combiner, and the Atacama
Large Millimeter Array (ALMA).
First,~\citet{gravity18} \citep[see also][]{gravity23}  observed a series of polarized infrared flares. The QU pattern traces a single loop during the observed astrometric orbital period. The authors show that this pattern is consistent with a hot spot orbiting at a radius close to the ISCO of a non-spinning black hole.
Second,~\citet{wielgus22} observed a QU loop with ALMA {at millimeter wavelengths, following an X-ray flare reported by Chandra \citep{wielgus22_generalLC}}. The authors show that the data are consistent with a hot spot orbiting at a radius about two times the ISCO of a non-spinning black hole, with the QU loop period interpreted as the Keplerian period of the hot spot. The hot spot interpretation is not unique though: the EVPA swings have been interpreted by \citet{yusefzadeh07} not in terms of an orbiting hot spot, but rather within the framework of an ejected expanding blob of plasma. This alternative model has recently been discussed by~\citet{michail23}.

In this article, we investigate the polarized synchrotron radiation emitted by orbiting hot spots. In this context, the orientation of the magnetic field has a crucial impact on the observables. Indeed, the electric vector (the orientation on sky of which is encoded in the
QU loop) is {oriented along the cross product} 
$\mathbf{K} \times \mathbf{B}$, where $\mathbf{K}$ is the photon's direction of emission and $\mathbf{B}$ is the magnetic field
vector, both expressed in the comoving frame of the emitter.
There is a growing body of evidence that the magnetic field
in the close surroundings of Sgr~A* is rather ordered, dynamically important (i.e. the plasma dynamics is sensitive to the
magnetic field), with a dominant poloidal component (i.e. in a plane orthogonal to the equatorial plane of the black hole). The hot spot modeling of infrared data performed by~\citet{gravity18,jimenez20} favors a strong poloidal field. The QU loop observed by \citet{wielgus22} favors a vertical field, while the persistence of the rotation measure, the sign of the circular polarization, and the magnitude
of the linear polarization fraction all favor a structured
magnetic field of persistent topology {\citep[see also][]{wielgus2023}}. The analysis of~\citet{michail23} favors a magnetic field orientation aligned with the angular momentum vector of the accretion flow, that is, vertical for an accretion flow centered on the equatorial plane of the black hole.
{The analysis of the spatially resolved event horizon scale images of Sgr~A* obtained by the Event Horizon Telescope \citep[EHT; ][]{EHT_SgraP1,EHT_SgraP3} further supports the magnetically arrested disk (MAD) accretion flow model interpretation \citep{EHT_SgraP5}, characterized by dynamically important magnetic fields with a strong vertical component near the event horizon \citep{narayan03}.} Furthermore, ordered magnetic fields in the compact region around Sgr~A* were revealed by pre-EHT very long baseline interferometry polarimetric observations \citep{Johnson2015}.
%The picture of an ordered, dynamically
%important, poloidal magnetic field is consistent with a
%magnetically arrested disk~\citep{narayan03} \mw{accretion mode}.

We aim to study the properties of QU loops
associated with hot spots around black holes. Such investigations are the subject of recent intense theoretical
efforts~\citep{jimenez20,gelles21,Narayan2021,vos22,gravity23,Najafi2023}.
%\textbf{Report main features emerging.}
Here we intend to contribute to this emerging topic by mainly focusing on the impact of special relativity on the observables.
We develop a thorough analysis of QU loops in Minkowski spacetime and show that these flat spacetime loops share the main features of their general-relativistic counterparts, demonstrating that QU loops are strongly affected by the relativistic velocities of their emitter, and the associated special-relativistic light aberration. 
%\mw{but they are mostly due to anisotropy of synchrotron for low inclinations, no?, "...primarily affected by the synchrotron emission anisotropy, the relativistic..."? } \fv{Yes the anisotropy of synchrotron strongly affects the flux evolution, thus the QU loops, but I think that this was already known to people (at least kind of known). The really new feature we add is the importance of aberration, that's why I insist on it here. Maybe my wording is incorrect: by "primarily" I mean "it matters a lot", I don't mean "it's the only important parameter".} \mw{ok, I'm fine with this, just consider if you prefer "primarily" or just "strongly". One other suggestion, maybe "...associated special-relativistic light aberration."?}
We also develop an analytical understanding of the properties of these QU loops. We then compute QU loops in the Schwarzschild spacetime, comparing them to their Minkowski counterparts and to the relevant
literature. The main aim of this article is to elucidate which aspects of these observable patterns are due to special-relativistic, and due to general-relativistic effects.

The paper is organized as follows. Section~\ref{sec:model} describes
our hot-spot model. Section~\ref{sec:introQU} introduces in detail
the topic of QU loops and all the necessary concepts.
Section~\ref{sec:Minko} is the main section of
the article and is dedicated to the properties of Minkowski QU loops.
Section~\ref{sec:Sch} describes Schwarzschild QU loops, and
section~\ref{sec:conclu} gives our conclusions and perspectives.

% Papers to cite:
% \begin{itemize}
% \item QUloop: All papers in my QUloop directroy + GRAVITY18
% %\item Hotspot: Broderick \& Loeb, Hamaus+09
% %\item Plasmoid and recon: see Nico's paper biblio
% %\item IR polar Obs: Eckart+06 Polarimetry of..., Trippe+07 A polarized infrared..., Nishiyama+09, Zamaninasab+10? Shahzamanian+14?
% %\item Radio polar Obs: Marrone+06, Wielgus+22, Michail+23 Polarized signatures..., 
% %\item Non-Kerr objects: TBC
% \end{itemize}

% To be discussed
% \begin{itemize}
% \item 1. Bfield config; 2. Question: can hot spots be probes of spacetime? 3. Non-Kerr spacetimes and RZ; 4. polarized GR codes for QUloops; 5. Contents.
% \item RZ parameters value (in model section)
% \item Bfield config: simple picture of 10 years ago (toroidal inside disk coz frozen in plasma), poloidal outside of disk (along the jet direction?); picture stemming from Benoit,Ileyk+ and Bart+ simus.
% %\item Bfield constrains for Sgr, rather ordered vertical field close to EH: see Gold+17, EHT papers, GRAVITY18, Maciek22, Michail23
% \item See Meyer+06 ring+hotspot model? See Ricarte+22 general discussion on polarized observables?
% \end{itemize}

%=========================================
%=========================================
\section{Modeling hot spot observables}
\label{sec:model}
%=========================================
%=========================================

In this section we present our model of a rotating hot spot
around a compact object. We discuss the spacetime geometry, the
shape, physical characteristics and emission of the hot spot,
and the radiative transfer integration by means of relativistic ray tracing. {We consider physically motivated values of the model parameters. For a more extensive discussion of the impact of the individual parameters on the QU loop patterns see \citet{vos22}.  }

\subsection{Spacetime geometry}

The main aim of this article is to discuss the respective influence of special- and general-relativistic effects on the polarized signatures associated with orbiting hot spots. To that end, we perform calculations in Minkowski and Schwarzschild spacetimes.
%to Minkowski spacetime and discuss them
%in comparison with their properties in the Schwarzchild spacetime. 
%\mw{I would rephrase the main aim into characterizing what's non-relativistic, what's SR, what's GR in the QU loops. I think generally somewhere a caveat would be good to mention, that when we say that the features are explained with SR by the dynamics of the hot spot respectively of the spacetime curvature of course, we are not discussing WHY the dynamics are such and such (this is a bit where I find the narrative a bit awkward overall - because spacetime curvature does impact dynamics (it's such and such orbit because such and such curvature imply such and such geodesic, and this orbit may manifest through SR effects, Doppler, aberration, but may very well be used to diagnose the GR curvature) so the one may ask what are we actually doing, are we suggesting that there is no gravitating supermassive compact object in the GC? We should explain ourselves clearly. )} \fv{Is it sufficiently clear with the last sentence added in the Intro? I have also rephrased the first sentence of this section.}
We consider that the spacetime is
described in spherical coordinates $(t,r,\theta,\pp)$.
We assume that the spacetime is {static} and spherically symmetric,
meaning that we will not discuss any impact of the compact object's spin
in this article.
The metric line element thus reads
\begin{align}
  \dd s^2 &= g_{tt} \,\dd t^2 + g_{rr} \, \dd r^2 + g_{\theta\theta} \,\dd \theta^2 + g_{\pp\pp} \,\dd \pp^2\\ \nn
          &=g_{tt} \,\dd t^2 + g_{rr} \, \dd r^2 + r^2 \left( \dd \theta^2 + \sin^2 \theta \,\dd \pp^2\right) \\ \nn
\end{align}
where $g_{\mu\nu} = \boldsymbol{\partial_\mu} \cdot \boldsymbol{\partial_\nu}$
are the metric coefficients, that can be expressed as the dot products
between the natural basis vectors associated with the spherical coordinates,
$\boldsymbol{\partial_\mu}$. The two static spacetimes (Minkowski,
Schwarzschild)
that we will consider
are thus fully defined by their metric coefficients $g_{tt}$ and $g_{rr}$.
%In the following, we need to specify not only the spacetime metric
%coefficients $g_{tt}$ and $g_{rr}$, but we also need to express
%the Keplerian 4-velocity for circular orbits in the equatorial plane.
%The motion of the hot spot indeed has a crucial impact on the observable.

\subsubsection{Minkowski spacetime}

As a flat manifold, Minkowski spacetime has little a priori relevance for interpreting data originating from the close environment of a supermassive black hole. However, by studying QU loops in this context we aim at revealing key observable features to discuss and interpret and, in particular, to be able to tell which aspects are specific to spacetime curvature, and which aspects are already present in a flat spacetime. The Minkowski metric is defined with
\be
g_{tt} = -1, \quad g_{rr} = 1.
\ee

\subsubsection{Schwarzschild spacetime}

The Schwarzschild metric in Schwarzschild coordinates is
\be
\label{eq:Sch}
g_{tt} = -\left(1 - \frac{r_S}{r}\right), \quad g_{rr} = \left(1 - \frac{r_S}{r}\right)^{-1},
\ee
where $r_S = 2M$ is the location of the Schwarzschild event horizon. 
The Jebsen-Birkhoff\footnote{The famous Birkhoff's theorem of general relativity, published by Birkhoff in 1923, was first published two years before by the Norwegian physicist Jebsen~\citep[see][for an historical account]{johansen05}.} theorem~\citep{jebsen21,birkhoff23} ensures that the Schwarzschild geometry uniquely describes the
spacetime outside of any spherically symmetric object in vacuum. We note that we use throughout this article a system of units where the gravitational constant $G$
and light speed $c$ have unit values, so that the gravitational radius, $GM/c^2$, is simply equal to $M$.

% \subsubsection{Rezzolla-Zhidenko spacetime}

% In order to assess the impact of the spaceime geometry on
% the polarized observables, we will consider a parametrized
% departure to the standard static black hole metric.
% Such a geometry has been proposed by~\citet{rezzolla14},
% hereafter refered to as the RZ spacetime. The metric reads
% \begin{align}
%   g_{tt} &= -N^2(r), \\ \nn
%   g_{rr} &= \frac{1}{N^2(r)}, \\ \nn
%   N^2(x) &= x\, \left[ 1 + a_1 \left( 1 - x\right)^3\right], \quad x = 1 - \frac{2M}{r},
% \end{align}
% where $a_1$ is a parameter describing the departure from Schwarzschild
% geometry. It is obvious to check that $a_1=0$ coincides with the
% Schwarzschild metric of Eq.~\ref{eq:Sch}. The expression above is
% considerably simplified compared to the most generic parametrized
% metric presented in~\citet{rezzolla14}. We have decided to
% follow~\citet{bauer22} and consider only the first non-constrained
% parameter, $a_1$, and we refer to their discussion for further details.

\subsection{Hot spot geometry and physical quantities}

%We consider a hot spot made of a coordinate sphere following a timelike
%circular geodesic at a coordinate radius $r_0$
%%in the equatorial plane. That is, the center of the sphere
%follows a Keplerian orbit at $r=r_0$ in the equatorial plane $\theta=\pi/2$,
%and the hot spot is made
%of radiating electrons located within a certain coordinate radial distance from
%this center.
Let us consider two spatial positions $P_0(x_0,y_0,z_0)$ and $P(x,y,z)$,
where the Cartesian coordinates are related to the spherical ones by
means of standard Euclidean formulas. We define the coordinate distance between 
$P_0$ and $P$ as the Euclidean distance defined by 
$d^2 = (x-x_0)^2 + (y-y_0)^2 + (z-z_0)^2$.
The center of our hot spot is located at a constant radius $r_0$,
in the equatorial plane $\theta_0=\pi/2$, with a varying azimuthal
angle $\pp_0$.
%\mw{what's a "coordinate sphere"? I'd just say something like "spherical distribution of matter centered at radius r0" and then equations. Also this $\sigma_r$ is peculiar, missing sqrt(2) to be a gaussian std. Maybe just write somewhere that this choice of $\sigma_r = r_g$ means FWHM of 2.773 (which is physically the approximate diameter of the hot spot), or maybe just use a=FWHM in eq 4? so the exponential would have $-4\ln 2 (r-r_0)^2 / a^2$. Bonus, if you move to notation a=FWHM it frees $\sigma$ symbol to mean magnetization:)  }
%\fv{My bad I just forgot the factor 2. Sigma is indeed the std dev. I have added a definition of the "coordinate distance" above, and use it in the profiles below, the previous formulas were actually wrong.}
The hot spot is described by specifying the profiles of the electron number density
$n_\mathrm{e}$, temperature of the electrons $T_\mathrm{e}$, %\mw{minor issue, you use $T_\mathrm{e0}$ in Tab 1 and $4T_\mathrm{e0}$ in eq. 4, homogenize this, i like mathrm more} 
as well as the magnitude
and direction of the ambient magnetic field $B$.
All these quantities are measured in the
rest frame of the orbiting hot spot, that we will hereafter refer to as the emitter's frame.
%The motion followed by
%the hot spot and the direction of the magnetic field are discussed in sections~\ref{sec:motion} and~\ref{sec:Bfield},
%while the emission properties are presented in~\ref{sec:radtrans}.
We assume the following profiles for the physical quantities
\begin{align}
  \label{eq:profiles}
  n_\mathrm{e} &= n_\mathrm{e0}\,\mathrm{exp} \left(- \frac{d^2}{2 \sigma_r^2} \right), \\ \nn
  T_\mathrm{e} &= T_\mathrm{e0}\,\mathrm{exp} \left(- \frac{d^2}{2 \sigma_r^2} \right), \\ \nn
  \frac{B^2}{4 \pi} &= \eta \, m_p c^2 \, n_\mathrm{e}, \\ \nn
\end{align}
where $d^2$ is the squared coordinate distance (as defined at the beginning
of this section) to the center of the hot
spot $(x_0=r_0 \cos \pp_0,y_0=r_0 \sin \pp_0,z_0=0)$, $n_\mathrm{e0}$ and $T_\mathrm{e0}$ are the density and temperature at the center
of the hot spot,
$\sigma_r$ is the Gaussian standard deviation, which is related to the
full width at half maximum, that is, to the effective diameter $D_{\rm hs}$ of the
hot spot, by $D_{\rm hs} \approx 2.35\,\sigma_r$,
$\eta$ is the magnetization parameter, $m_p$ is the proton rest mass,
and we assume a constant ratio between the particle rest-mass and magnetic energy densities.
%\mw{what does equipartition mean for these equations? $\eta = 1$? but this ignores kinetic energy of particles... I think you mean that $\eta$ is constant in the domain, but is this really equipartition? Isn't equipartition saying that the different types of energies are the same?}.
%\fv{Yes it's probably not perfect wording but the idea is to say that we consider the particle's rest mass energy on the one hand, the magnetic energy on the other hand, and we simply prescribe some constant ratio between the two. Indeed strictly speaking equipartition is $\eta=1$ only. Not sure how to phrase it better, I have the feeling that people will understand like that, but it could certainly be made more precise.} \mw{so isn't it better to just say (a bit longer but more precise) that "we assume a constant ratio between the magnetic and rest mass energy densities"?}
Furthermore, we assume that the hot spot is described by a spatial
Gaussian profile around its center, while its properties remain constant in time.

The values assumed for the parameters introduced so far are listed in Table~\ref{tab:properties}. 
We note that the magnetic field maximum magnitude ($B_0 = 140$~G, a consequence of the simple prescription given by the third line of Eq.~\ref{eq:profiles}) is rather high as compared to the typical value that can be derived from the synchrotron cooling time~\citep[see e.g.][]{aimar23a}, but the precise value anyway does not impact the results of this article.
We also note that we fix the magnetization
to $\eta=1$ corresponding to
a strongly magnetized flow, in agreement with hints that Sgr~A* is
likely a magnetically arrested flow~\citep[e.g.][]{gravity18, EHT_SgraP5, wielgus22}.
The central density and temperature are chosen to ensure a
near infrared maximum dereddened flux (that, is corrected from the strong extinction towards the Galactic center) of $\approx 10$~mJy at low inclination for a vertical
magnetic field. This value corresponds to a rather bright infrared
flare, see the percentiles of Sgr~A* dereddened flux distribution
provided in Table~1 of~\citet{gravity20_sebo}, and can be compared
first to the dereddened flux density of S2 that reaches $\approx 16$~mJy~\citep{gravity20_sebo}, and second to the brightest
infrared flare ever observed that reached $\approx 60$~mJy~\citep{do19}.
%\mw{is this 10 mJy arbitrary or is this some estimate in physical units of what GRAVITY is typically seeing during NIR flares? if the latter, that's a good thing to mention, possibly with a reference} 
We note that other configurations, with different magnetic field geometry, can lead to much higher fluxes that are not in agreement with observations. Nonetheless, we keep the central density and temperature fixed in order to ease the interpretation of the impact of the magnetic field gometry on the observables.
%enable the comparison between the polarized signatures between different model parameters.

\renewcommand{\arraystretch}{1.2}
\begin{table}[h!]
\centering
 \begin{tabular}{c c c }
 \hline
 Symbol &  Value & Property \\  %\vspace{0.5cm}                                 
 \hline
 $M$ & 4.3 $\times 10^{6}$~M$_\odot$ & compact object mass  \\
  %\hline                                                                       
   $D$ & 8.28~kpc & compact object distance \\
   $a$ & 0 & BH spin parameter \\
   % \hline
   $r_0$ & $8\,r_g$ & hot spot orbital radius  \\
   $\sigma_r$ & $r_g$ & hot spot Gaussian extension  \\
 $n_\mathrm{e0}$ & $2 \times 10^6\,\mathrm{cm}^{-3}$ & max number density of electrons  \\
 %\hline                                                                        
   $T_\mathrm{e0}$ & $10^{11}$~K & max electron temperature \\
   $B_0$ & $140$~G & max magnetic field \\
 %\hline                                                                        
   $\eta$ & 1 & magnetization \\
   $\kappa$ & 4 & index of $\kappa$ electron distribution \\
 %\hline                                                                        
 $i$ & $[90^\circ - 180^\circ]$ & inclination angle\\
  %\hline                                                                       
  % $PA$ & -70$^\circ$ & jet position angle east of north \\ %in the image \\   
  %\hline                                                                       
 $\lambda_\mathrm{obs}$  & $2.2\,\mu$m~& observing wavelength \\
   %\hline                                                                      
 $f$ & 200~$\mu$as & field of view \\
    %\hline                                                                     
 $N \times N$ & 128$\times$128 & image resolution \\
 [1ex]
 \hline
\end{tabular}
\caption{Parameters of our model. The mass and distance to Sgr~A* are
  taken from~\citet{gravity20,gravity21}. The orbital radius is close
  to that found by~\citet{gravity18,wielgus22}.
  The density and temperature are
  chosen to ensure a $2.2\,\mu$m dereddened flux of the order of $10$~mJy.
  The magnetic field is linked to the density through the
  assumption of Eq.~\ref{eq:profiles}. 
  It is still listed here for completeness. We remind that the inclination angle $i$ corresponds to the Boyer-Lindquist $\theta$ angle (illustrated in Fig.~\ref{fig:QUbasic}) of the observer. In the text, the complementary angle $\iota = \pi - i$ is often used.
  %The magnetization is high, which is in agreement with Sgr~A* being
  %a magnetically arrested flow~\citep[e.g.][]{gravity18,EHT_SgraP5,wielgus22}.
  }
\label{tab:properties}
\end{table} 

\subsection{Hot spot motion}
\label{sec:motion}

The hot spot center located at $r_0$ is assumed to follow a
circular timelike geodesic, that is, a Keplerian orbit of the
spacetime considered. Its 4-velocity thus reads
\be
\mathbf{u} =  u^t \left(\boldsymbol{\partial_t} + \Omega \, \boldsymbol{\partial_\pp} \right), \quad \Omega = \frac{u^\pp}{u^t}.
\ee 
The expressions of $u^t$ and $\Omega$ depend on the spacetime metric.
That of $\Omega$ is well known for the Schwarzschild spacetime expressed
in Schwarzschild coordinates,
$\Omega_\mathrm{Schwarzschild} = M^{1/2} \, r^{-3/2}$~\citep[e.g.][]{bardeen72}.
Given that this expression coincides with the Newtonian result, we use the
same expression in Minkowski, even though there is no reason for the hot spot to follow orbital motion in a flat spacetime (in the absence of a central massive object). Hence, we only consider Minkowski spacetime to determine what features of the observables are specific to a curved spacetime, and what features are already present in a flat geometry.
%Nonetheless, we consider Minkowski spacetime in this paper for pedagogical purposes.
%The rotation velocity of the RZ spacetime was
%derived e.g. in~\citet{cardenas19}.
With an expression for $\Omega$,
it is straightforward to derive that of $u^t$ by using the normalization
of the 4-velocity, $\mathbf{u} \cdot \mathbf{u} = -1$.
We finally obtain 
\begin{align}
  \label{eq:velo}
  u^t &= \sqrt{\frac{r}{r-M}}, \quad \Omega = M^{1/2} \, r^{-3/2}, \quad \mathrm{[Minkowski]} \\ \nn
  u^t &= \sqrt{\frac{r}{r-3M}}, \quad \Omega = M^{1/2} \, r^{-3/2}. \quad \mathrm{[Schwarzschild]} \\ \nn
  %u^t &= \frac{1}{\sqrt{N\left(N - r N' \right)}}, \quad\Omega = \sqrt{\frac{N N'}{r}}, \quad \mathrm{[RZ]} \\ \nn
\end{align}
%where $N'$ is the radial derivative of $N(r)$.

% Note that there is no reason that a hot spot in flat spacetime would orbit,
% in the absence of any central body. However, and only for the pedagogical
% reasons evoked above, we will consider that the Minkowski hot spot
% has the following 4-velocity
% \be
% \label{eq:4velM}
% \mathbf{u}_M = u^t_M \left(\boldsymbol{\partial_t} + \Omega_M \, \boldsymbol{\partial_\pp} \right), \quad \Omega_M = r^{-3/2},
% \ee
% where the subscript M refers to Minkowski, $\boldsymbol{\partial_\mu}$
% are the basis 4-vectors naturally associated with the Cartesian coordinates
% in Minkowski spacetime, and $\Omega_M$ is the rotation velocity of the
% hot spot. We have considered that $\Omega_M$ follows the Newtonian
% profile $r^{-3/2}$. The normalization of the 4-velocity, $\mathbf{u}_M \cdot \mathbf{u}_M = -1$, then allows to obtain
% \be
% u^t_M = \sqrt{\frac{r}{r-M}},
% \ee
% defined for $r>M$. We note that we use a system of units
% where the gravitational constant $G$
% and light speed $c$ have unit values, so that the gravitational
% radius, $GM/c^2$, is simply equal to $M$.

% The Keplerian 4-velocity in the Schwarzschild spacetime
% takes the same expression as in Minkowski,
% \be
% \mathbf{u}_S = u^t_S \left(\boldsymbol{\partial_t} + \Omega_S \, \boldsymbol{\partial_\pp} \right), \quad \Omega_S = r^{-3/2},
% \ee
% however the normalization of the 4-velocity leads to
% \be
% u^t_S = \sqrt{\frac{r}{r-3M}},
% \ee
% defined for $r>3M$, that is, above the Schwarzschild photon orbit
% located at $r=3M$.

\subsection{Magnetic field configuration}
\label{sec:Bfield}

We have so far only defined the magnitude of the magnetic field
through Eq.~\ref{eq:profiles}. We proceed to specify its direction,
hence we need to define a unit spacelike vector, normal to the hot spot
4-velocity given that the magnetic field vector lies in the rest space of the
emitter. 

We will consider only two different configurations: either vertical, or toroidal. 
These two configurations are inspired by two plausible magnetic configurations that could exist around Sgr~A*. Either the environment is weakly magnetized and the magnetic field lines will follow the motion of the matter swirling towards the black hole, in which case the magnetic field will be mostly toroidal (this would correspond to a SANE -- standard and normal evolution -- situation), or the environment is strongly magnetized and the magnetic field does not follow the motion of the matter, in which case it would have a strong vertical component like in MAD states.
Thus, we define
\begin{align}
  \mathbf{\bar{B}} &= (0,B^r,B^\theta,0), \quad \mathrm{[Vertical]} \\ \nn
  \mathbf{\bar{B}} &= (B^t,0,0,B^\pp), \quad \mathrm{[Toroidal]} \\ \nn
\end{align}
where the upper bar means that the vector is a unit vector,
with the constraints that
\be
\mathbf{\bar{B}} \cdot \mathbf{\bar{B}} = 1, \quad \mathbf{\bar{B}} \cdot \mathbf{u} = 0.
\ee
The second condition implies that the magnetic field $\mathbf{\bar{B}}$ lies in the local rest space of the emitter, that is, the space orthogonal to its 4-velocity. We are thus defining the magnetic field as measured by the emitter.
These conditions immediately lead to
\begin{align}
  \label{eq:Bdef}
  \mathbf{\bar{B}} &= \cos \theta \,\boldsymbol{e_r} - \sin \theta \,\boldsymbol{e_\theta} , \quad \mathrm{[Vertical]} \\ \nn
  \mathbf{\bar{B}} &= \frac{1}{\sqrt{-\left(g_{tt} + \Omega^2 g_{\pp\pp}\right)}} \left( \sqrt{-\frac{g_{\pp\pp}}{g_{tt}}} \, \Omega \, \boldsymbol{\partial_t} + \sqrt{-\frac{g_{tt}}{g_{\pp\pp}}} \,\boldsymbol{\partial_\pp}  \right), \! \! \quad \! \! \mathrm{[Toroidal]} \\ \nn
\end{align}
where $\Omega$ is the Keplerian rotation velocity defined in Eq.~\ref{eq:velo},
and we use the orthonormal basis associated to the natural coordinate
basis $\boldsymbol{\partial_\mu}$
\be 
  \label{eq:orthbasis}
  \boldsymbol{e_t} = \frac{\boldsymbol{\partial_t}}{\sqrt{-g_{tt}}}, \:
  \boldsymbol{e_r} = \frac{\boldsymbol{\partial_r}}{\sqrt{g_{rr}}}, \:
  \boldsymbol{e_\theta} = \frac{\boldsymbol{\partial_\theta}}{\sqrt{g_{\theta\theta}}}, \:
  \boldsymbol{e_\pp} = \frac{\boldsymbol{\partial_\pp}}{\sqrt{g_{\pp\pp}}}.
\ee
This basis coincides with the locally non-rotating frame~\citep{bardeen72} of the Schwarzschild spacetime.
Note that although the hot spot's center $r_0$ orbits in the
equatorial plane, the full hot spot is a 3-dimensional structure in
space and is not restricted to the equatorial plane. This is why the
magnetic field is defined for all $\theta$ and not only
for $\theta=\pi/2$.

\subsection{Radiative transfer}
\label{sec:radtrans}

The hot spot is assumed to emit synchrotron radiation, and the
emitting electrons are considered to follow a $\kappa$ distribution,
that is, a mix between a thermal core and a power-law tail. This
distribution is well adapted to simulate the state of electrons
locally accelerated (for instance through magnetic reconnection)
that radiate during Sgr~A* flares. {The $\kappa$ distribution is thus a more physical assumption, particularly for the infrared emission during a flare, than the thermal spectrum considered by \citet{wielgus22} and \citet{vos22}.} This distribution reads
\be
n_\mathrm{e}(\gamma) = N \,\gamma (\gamma^2 - 1)^{1/2} \left( 1+\
 \frac{\gamma-1}{\kappa \theta_\mathrm{e}}\right)^{-(\kappa+1)}
\ee
where $\gamma$ is the Lorentz factor of the electrons,
$N$ is a normalizing coefficient chosen such that the integral
of $n_\mathrm{e}(\gamma) $ over all $\gamma$ is equal to the total number density
of the hot spot, $\theta_\mathrm{e} = k T_\mathrm{e} / m_\mathrm{e} c^2$ is the dimensionless electron
temperature, with $k$ and $m_\mathrm{e}$ being the Boltzmann constant and
electron rest mass. We chose a parameter $\kappa = 4$. This translates
to an infrared spectral index $\alpha=0$ where $\nu F_\nu \propto \nu^\alpha$,
which is reasonable for bright flares~\citep{gillessen06}.
We utilize the emission, absorption, and Faraday rotation/conversion coefficients
for $\kappa$-synchrotron as derived by~\citet{marszewski21}. 
These coefficients have rather complicated and lengthy expressions that
we do not fully repeat here. However, it will be useful for forthcoming discussion to indicate that the emission coefficients for the various Stokes parameters are expressed as
\be
\label{eq:jnuPolar}
j_\nu \propto \frac{n_\mathrm{e} e^2 \nu_c}{c} X_\kappa^{-(\kappa-2)/2}  \, \sin \theta_B, \hspace{0.1cm} \left\{
    \begin{array}{ll}
        \propto \sin^2 \theta_B, & \\
        \propto \nu^{-1}, &
    \end{array}
\right.
[\mbox{for} \hspace{0.1cm} \kappa=4]
\ee
where $X_\kappa = \nu [\nu_c (\theta_\mathrm{e} \kappa)^2 \sin \theta_B]^{-1}$, $\nu_c$ is the cyclotron frequency, and $\theta_B$ is the angle between the magnetic field direction and the direction of emission. The proportionality factor in the above expressions depends on $\kappa$ and on the particular Stokes parameter that is considered. 
This expression coincides with the so-called high-frequency emission coefficient reported in Eq.~44 of \citet{marszewski21}, which applies for our typical conditions.
%\mw{\st{It will reveal crucial for future discussion to note the strong directionality of this expression, that appears clearly in the $\sin \theta_B$ term. }}
The strong directional dependence of this expression, evident in the $\sin \theta_B$ term, will be crucial for the forthcoming discussion. We note that for $\kappa=4$, the expression behaves as $\sin^2 \theta_B$, so that it cancels in the direction of emission along the magnetic field lines, and reaches maximum in the direction normal to the magnetic field. We also note that the frequency dependence of the emission coefficient follows $\nu^{-1}$. {While the Faraday effects are generally negligible for modeling infrared flares, they become important at millimeter wavelengths, for which significant Faraday rotation is most likely associated with the compact emission region, contributing non-trivially to the observed complex linear polarization \citep{wielgus2023}. }

%\textbf{Give (maybe simplified?) expressions, important $\sin \theta_B$
%dependence to be discussed.}

\subsection{Polarized ray tracing}

We compute the polarized flux emanating from the orbiting hot spot
by using the \gy code~\citep{vincent11,aimar23b}. We consider an
observer located at a distance $D = 8.28$~kpc~\citep{gravity21}.
The compact object's mass is fixed to
$M = 4.3 \times 10^6\,M_\odot$~\citep{gravity20}.
The inclination (corresponding to the spherical coordinate
$\theta$) is varied in $[90^\circ, 180^\circ]$, with $90^\circ$
corresponding to an edge-on view, and $180^\circ$ to a face-on view.
This range encompasses the best-fit inclination
for Sgr~A* of $\approx 160^\circ$
derived
by~\citet{gravity18,wielgus22}. Inclinations higher than $90^\circ$ recover a clockwise motion on sky of the hot spot, consistent with observations.

Null geodesics are traced backwards from the observer's screen towards the
hot spot, and the full polarized radiative transfer is solved. We account for the finite velocity of light (so-called "slow-light" paradigm).
The final product of the computation
is a set of maps of the specific Stokes parameters
$(I_\nu,Q_\nu,U_\nu)$, introduced in section~\ref{sec:Stokes},  for the various orbital phases of the hot spot. 
%see section~\ref{sec:Stokes},
We discard Stokes V in this article, although it is computed.
We always consider a resolution of $N \times N = 128 \times 128$ pixels,
and a field of view of $f = 200 \,\mu$as. The observing wavelength
is set to $\lambda_\mathrm{obs} = 2.2 \,\mu$m, coinciding
with that of the GRAVITY instrument. All parameters discussed in this section are listed in
Table~\ref{tab:properties}.

\section{Polarization signature of hot spots}
\label{sec:introQU}

Before turning to the detailed properties of QU loops
that will be discussed in the context of Minkowski spacetime
in the next section, in this section we introduce all relevant material for the following discussions. We will define the Stokes Q and U parameters, the electric vector position angle,
and intuitively introduce the concept of QU loops associated
with orbiting hot spots.

\subsection{Stokes Q and U parameters, observed EVPA}
\label{sec:Stokes}

%We briefly remind here the definitions of
%the Stokes Q and U parameters, that are among the core observables
%associated to orbiting hot spots.

We consider a linearly polarized wave incident on
the observer's screen. This is a simplification in the sense
that synchrotron radiation is mostly linearly polarized but
has non-zero circular polarization. Given that in this
article we will never discuss circular polarization, we
only introduce here the linearly polarized part of the radiation,
encoded in the Stokes Q and U parameters. Note that our
ray-tracing calculations consider the full synchrotron
radiative transfer, with also non-zero Stokes V.

\begin{figure}[htbp] 
\centering 
\includegraphics[width=0.4\textwidth]{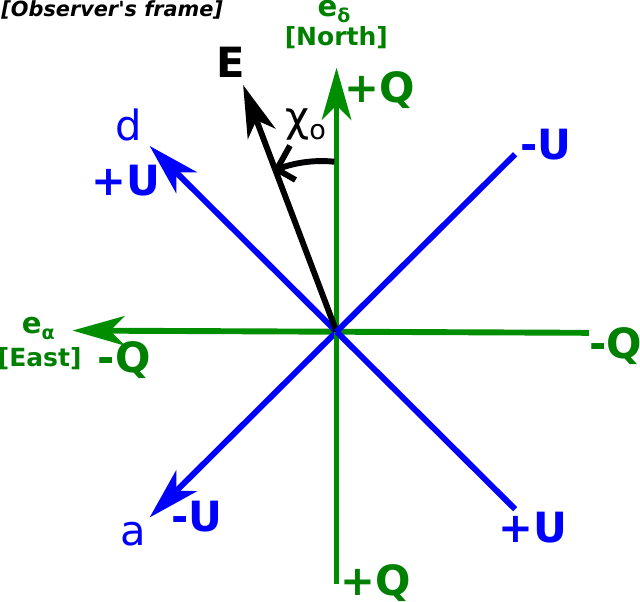} 
\caption{Electric field, observed EVPA and Stokes Q and U. All quantities are defined in the observer's frame, as measured by the distant observer.
  The observed electric field associated
  to the wave received at the observer's screen is the black arrow,
  with a position angle East of North corresponding to the observed Electric Vector
  Position Angle, or observed EVPA. For a fully linearly polarized wave,
  there is a bijection (up to a sign ambiguity)
  between providing the electric vector magnitude and direction on
  screen, and the pair of Stokes parameters $(Q,U)$.
  The electric vector magnitude is given by $\sqrt{Q^2+U^2}$,
  while its orientation 
  follows
  $\chi_\mathrm{o} = {1}/{2} \,\mathrm{atan2}(Q,U)$,
  see Eq.~\ref{eq:EVPAdef}. It is easy to check from the
  definitions of Eqs.~\ref{eq:QUdef} that the North-South
  and East-West directions coincide with positive and negative
  Stokes Q (and zero Stokes U), respectively, while the
  diagonals correspond to positive and negative Stokes U (and
  zero Stokes Q), respectively.
} 
\label{fig:Stokes} 
\end{figure}

The electric field describing the incident wave on the
observer's screen is
\be
\label{eq:EE}
\mathbf{E} = E \left(\cos \chi_\mathrm{o} \,\boldsymbol{e_\delta} + \sin \chi_\mathrm{o} \,\boldsymbol{e_\alpha} \right) 
\ee
where $(\boldsymbol{e_\alpha},\boldsymbol{e_\delta})$ are the
unit vectors in the plane of the screen of the observer,
pointing towards the East and North  directions respectively,
see Fig.~\ref{fig:Stokes} for an illustration.
The angle $\chi_\mathrm{o}$, called the observed electric vector position angle (EVPA)
lies East of North from the North direction. The index $o$ is there to remind that this angle is defined in the observer's frame, hence the name of observed EVPA. We will introduce below an emission EVPA, defined in the emitter's frame.

The linear polarization information is encoded in the observed EVPA,
but this angle is not directly observable.
It is useful to introduce the following Stokes parameters
\begin{align}
  \label{eq:QUdef}
  Q &= E_\delta^2 - E_\alpha^2, \\ \nn
  U &= E_d^2 - E_a^2, \\ \nn
\end{align}
where the various $E_i$ represent the coordinate of
the electric vector along the corresponding directions
illustrated in Fig.~\ref{fig:Stokes}.
These are observable quantities, equal to differences
of intensities along specific directions on sky.
Equations~\ref{eq:EE} and~\ref{eq:QUdef} immediately lead to
\be
Q = E^2 \left( \cos^2 \chi_\mathrm{o} - \sin^2 \chi_\mathrm{o} \right) = I\, \cos 2\chi_\mathrm{o},
\ee
where $I = E^2$ is the total intensity, or Stokes I parameter.
Expressing the electric vector in the basis $(\mathbf{e_a},
\mathbf{e_d})$ associated to the directions $(a,d)$
rotated by $45^\circ$ with respect to $(\alpha,\delta)$,
see Fig.~\ref{fig:Stokes}, it is straightforward to obtain
\be
U = I\, \sin 2\chi_\mathrm{o},
\ee
so that the observed EVPA is simply obtained by
\be
\label{eq:EVPAdef}
\chi_\mathrm{o} = \frac{1}{2} \,\mathrm{atan2} \left(Q,U\right).
\ee
This angle lies in the range
\be
\chi_\mathrm{o} \in [-\pi/2, \pi/2],
\ee
and is defined modulo $\pi$, given that it only encodes
the direction of oscillation of the electric field.

\subsection{Emitter's and observer's bases, emission EVPA}

\begin{figure}[htbp] 
\centering  
\includegraphics[width=0.4\textwidth]{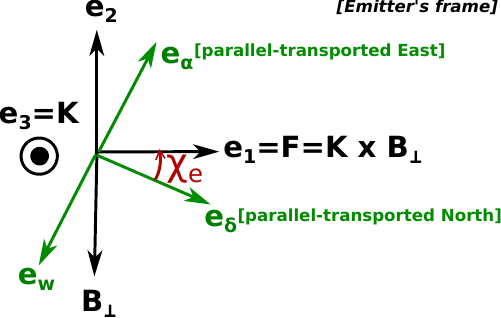} 
\caption{Emitter's and observer's polarization bases. All vectors discussed here are expressed in the emitter's frame.
The direction of
  emission is $\mathbf{K}$, while $\mathbf{B_\perp}$ is the ambient magnetic
  field projected normal to $\mathbf{K}$.
  The emitter's frame polarization vector reads $\mathbf{F} = \mathbf{K} \times \mathbf{B} = \mathbf{K} \times \mathbf{B_\perp}$.
  The vectors $(\mathbf{e_1},\mathbf{e_2},\mathbf{e_3})=(\mathbf{F},-\mathbf{B_\perp},\mathbf{K})$ form the emitter's
  orthogonal basis, naturally adapted for expressing synchrotron radiative transfer.
  The polarization basis of the observer
  $(\boldsymbol{e_\alpha}, \boldsymbol{e_\delta})$,
  corresponding to unit vectors in the East and North directions,  
  has been
  parallel transported to the emitter's frame.
  The vector $\mathbf{e_w}= -\boldsymbol{e_\alpha}$ is along the West direction, such that $(\mathbf{e_w},\boldsymbol{e_\delta},\mathbf{K})$ forms the observer's orthogonal triad.
  The emission EVPA is the
  angle $\chi_\mathrm{e} = (\boldsymbol{e_\delta},\mathbf{F})$ evaluated East of North, lying in between the observer's and emitter's bases.
  It is expressed by Eq.~\ref{eq:chi}.
} 
\label{fig:SynchPolar}
\end{figure}

The natural basis for expressing synchrotron emission in the emitter's frame is the orthogonal triad made of the following three vectors, all defined in the emitter's frame:
\begin{itemize}
\item the direction of photon emission $\mathbf{K}$ measured by the emitter, 
%that is, the projection of the null tangent 4-vector to the photon's geodesic $\mathbf{k}$ normal to the emitter's 4-velocity $\mathbf{u}$,  
\item the magnetic field vector $\mathbf{B_\perp}$ projected orthogonally to $\mathbf{K}$, 
\item and the emitter's frame polarization vector $\mathbf{F}$, which reads
\be
\mathbf{F} = \mathbf{K} \times \mathbf{B}.
\ee
\end{itemize}
We call these vectors $(\mathbf{e_1},\mathbf{e_2},\mathbf{e_3})=(\mathbf{F},-\mathbf{B_\perp},\mathbf{K})$, and refer to them as the emitter's polarization basis.
They are illustrated by the black vectors in Fig.~\ref{fig:SynchPolar}. It is in this emitter's basis that the polarized synchrotron radiative transfer coefficient are given.

However, the observable Stokes parameters are defined in the observer's 
polarization basis, 
$(\boldsymbol{e_\alpha}, \boldsymbol{e_\delta})$,
corresponding to the unit vectors in the 
East and North directions on 
the observer's sky. We thus need to integrate the polarized radiative transfer equation in this observer-related basis, and thus transform from the emitter's basis to the observer's basis. For doing this, we need
to parallel-transport the $(\boldsymbol{e_\alpha}, \boldsymbol{e_\delta})$ basis from the observer to the emitter along the photon's geodesic. The resulting vectors, parallel-transported
to the emitter's frame, are illustrated by the green vectors in Fig.~\ref{fig:SynchPolar}. 
The angle $\chi_\mathrm{e}$ between the parallel-transported North direction
and the polarization vector $\mathbf{F}$ allows to rotate between the synchrotron-adapted emitter's basis and the observer's basis. We call this angle the emission EVPA, hence the index $e$ in our notation. This wording reminds that this angle is expressed in the emitter's basis, and allows to make an explicit difference with the observed EVPA, $\chi_\mathrm{o}$, introduced above. There is in general no equality between $\chi_\mathrm{e}$ and $\chi_\mathrm{o}$, for the simple reason that $\chi_\mathrm{e}$ evolves along the geodesic as radiative transfer equations are integrated in the region containing plasma. However, for our setup consisting of a very compact emission region with nearly homogeneous conditions of motion and magnetic field, the emission and observed EVPA are very nearly equal. The distinction that we introduced between $\chi_\mathrm{e}$ and $\chi_\mathrm{o}$ is thus not important for our results (and we will often simply refer to the EVPA, without precision), but we consider that it is still important to make the distinction. 

The emission EVPA can be easily computed in the emitter's frame, from the projections of the vector $\mathbf{B_\perp}$
on the parallel-transported observer's polarization basis axes:
\be
\label{eq:chi}
\chi_\mathrm{e} = \frac{\pi}{2} - \mathrm{atan2} \left( \mathbf{B_\perp} \cdot \mathbf{e_w}, \mathbf{B_\perp} \cdot \boldsymbol{e_\delta} \right),
\ee
where $\mathbf{e_w} = -\boldsymbol{e_\alpha}$ is the unit vector
in the West direction, parallel transported to the emitter.
%, and $\mathbf{B_\perp}$ is the projection of the
%magnetic field $\mathbf{B}$ orthogonal to the emission direction
%$\mathbf{K}$.
We note that $\mathbf{B_\perp}$ is not a unit vector in general,
contrary to $\mathbf{e_w}$ and $\boldsymbol{e_\delta}$,
but this does not change the result of the atan2 function in Eq.~\ref{eq:chi}.
The emission EVPA is a crucial quantity for integrating the polarized radiative transfer.
We refer to~\citet{aimar23b} for details.

\subsection{Newtonian QU loops}  
\label{sec:Newton}

\begin{figure*}[htbp] 
\centering  
\includegraphics[width=0.8\textwidth]{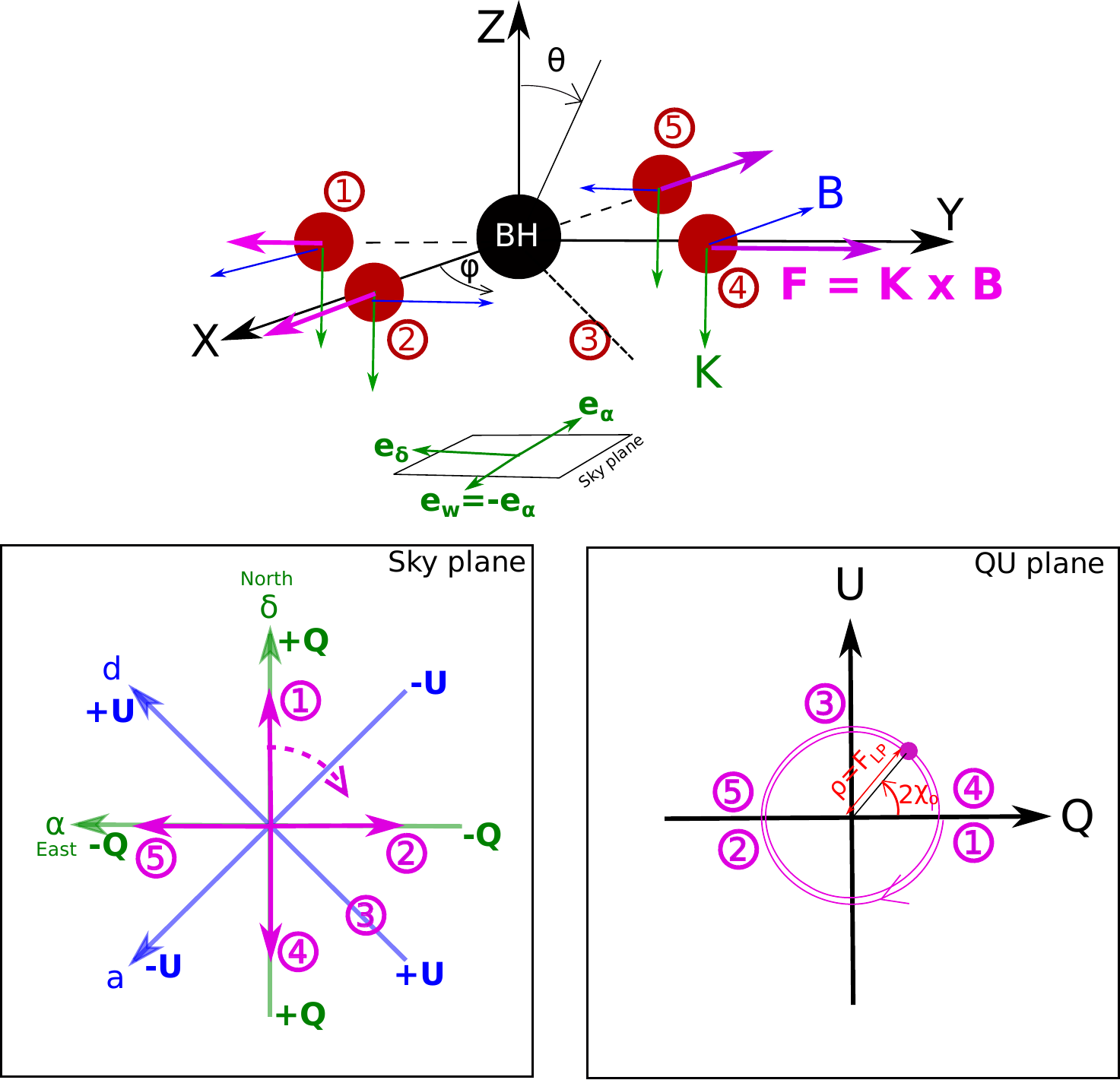} 
\caption{QU loop illustration in a non-relativistic context.
  \textbf{Top panel:}
  The black hole is represented by the
  black disk. The hot spot (red disk) orbits in the equatorial $XY$ plane
  around the black hole (black disk).
  The $Z$ axis is normal to the equatorial plane. We consider an observer
  looking face-on at the black hole, located towards the negative $Z$ axis.
  The North direction of the observer's screen is assumed to lie along
  the $-Y$ axis.
  The $\theta$ and $\pp$ angles of the spherical coordinates are
  represented. The hot spot rotates in the positive $\pp$ direction.
  The green vector $\mathbf{K}$ represents the direction of emission
  of the photon (we discard any relativistic effect here), the blue
  vector $\mathbf{B}$ is the magnetic field, assumed toroidal.
  The polarization vector $\mathbf{F} = \mathbf{K} \times \mathbf{B}$
  is shown in pink. Successive positions of the hot spot are labeled from
  1 to 5.
  \textbf{Bottom-left panel:} Rotation of the polarization vector on the
  sky plane of the observer, with the Stokes directions of
  Fig.~\ref{fig:Stokes} overlaid. \textbf{Bottom-right panel:} The associated
  QU plane and QU loops. The polar coordinates in this plane are $(\rho = F_{\mathrm{LP}} = \sqrt{Q^2 + U^2}, \phi = 2 \chi_\mathrm{o})$, where $F_{\mathrm{LP}}$ is the
  linearly polarized flux, and $\chi_\mathrm{o}$ is the observed EVPA. }  
\label{fig:QUbasic}
\end{figure*}

Let us consider a hot spot orbiting around a black hole, with a toroidal
ambient magnetic field, observed face-on by an infinitely distant observer,
as illustrated in Fig.~\ref{fig:QUbasic}.
Let us for the time being not consider any (special or general)
relativistic effect (that is, no lensing, no aberration, no relativistic Doppler or beaming effects). 
%\mw{mention specify what do you mean by that. No Doppler beaming, no relativistic aberration?}
The radiation is emitted in the vertical direction
along the vector $\mathbf{K}$.
It is easy to visualize that one complete rotation of the hot spot
will lead to a complete rotation of the 
polarization vector $\mathbf{F}$ in the plane of the sky,
as illustrated in Fig.~\ref{fig:QUbasic}. The bottom-right
panel of this figure shows that this leads to a double loop
in the QU plane. 
%\mw{[I wanted to add some comment about how these loops are essentially really basic, and more about symmetry than SR/GR, see if you agree and like it, modify or delete if you don't] 
Hence, at the most basic level, QU loops are a non-relativistic feature, simply a manifestation of an axisymmetric structure of the observed system.%}

If we consider the same setup as described above, but now
take a vertical magnetic field, our non-relativistic point of view
leads to concluding that the polarization vector would be
consistently zero ($\mathbf{K}$ and $\mathbf{B}$ being parallel)
as the hot spot rotates, leading to no QU loop.
As we will see in the next section, adding only special relativistic
effects (that is, still no light bending) allows to recover
QU loops in all cases, including for a face-on observer
with an ambient vertical magnetic field.
% The morphology of QU loops is very rich and complex,
% as we will see in the next section, and as was already
% investigated by several authors~\citep{jimenez20,gelles21,vos22}.
% However, note that in the simple discussion above, we have
% discarded all relativistic effects, and in particular special
% relativistic aberration effects. We will see in the next
% section that these effects are key for understanding the
% realistic signal.

%=========================================
%=========================================
\section{QU loops in Minkowski spacetime}
\label{sec:Minko}
%=========================================
%=========================================

In this section we derive an analytical understanding
of QU loops in Minkowski spacetime, and in particular we clarify
in what cases the rotating hot spot generates one or two loops
in the QU plane. Using Minkowski spacetime is helpful in order to gain insight in a simplified framework, without accounting for the light bending occurring in a curved spacetime. A non-intuitive conclusion of this section
is that all features of QU loops discussed in the literature in
the Schwarzschild or Kerr contexts are actually already present
in Minkowski. The crucial advantage of the flat geometry is that
exact analytical formulas can be derived to explain the QU loops.
The next three subsections are devoted to deriving an analytical
expression of the evolution of the emission EVPA depending on whether
the magnetic field is vertical or toroidal. This analytical model
is then compared to numerical simulations, which additionally constitutes a 
test of our polarized ray-tracing code.

\subsection{Direction of emission and aberration}

%\fv{I have added this subsection that does not depend on the orientation on B, to try to focus more clearly on the importance of the aberration effect. Hopefully this is a bit clearer?}

Let us consider a hot spot orbiting in Minkowski spacetime.
For the time being we do not specify the magnetic field orientation and only focus on the direction of emission in the emitter's frame.

The Minkowski 4-velocity of the emitter (that is, of the hot spot), defined in Eq.~\ref{eq:velo}, reads
\be
\label{eq:uuM}
\mathbf{u} = A \left(\boldsymbol{e_t} + r_0^{-1/2} \, \boldsymbol{e_\pp} \right),\quad A=\sqrt{\frac{r_0}{r_0-M}},
\ee
where we replaced the natural basis vectors $\boldsymbol{\partial_\mu}$
by the orthonormal basis vectors, using Eq.~\ref{eq:orthbasis}.

Let us consider an observer with an inclination $90^\circ \leq i \leq 180^\circ$.
We call $\iota = \pi - i$, which thus lies between $0$ and $90^\circ$.
The observer is assumed to be located at $\pp = -\pi/2$, that is,
in the $YZ$ plane (see Fig.~\ref{fig:QUbasic}).
The 4-vector tangent to the photon geodesic at emission reads
\begin{align}
\label{eq:k4vec}
  \mathbf{k} &= \mathbf{e_t} + \cos \iota \,\boldsymbol{e_\theta} - \sin \iota \, \mathbf{e_Y} \\ \nn
  &= \mathbf{e_t} - \sin \iota\, \sin \pp \,\mathbf{e_r} + \cos \iota \,\boldsymbol{e_\theta} - \sin \iota \, \cos \pp \,\boldsymbol{e_\pp}, \\ \nn
\end{align}
where $\mathbf{e_Y} = \sin \pp \, \mathbf{e_r} + \cos \pp \, \boldsymbol{e_\pp} $ is the unit vector along the $Y$
axis illustrated in Fig.~\ref{fig:QUbasic}. The vector $\mathbf{k}$
is clearly
a null vector of the Minkowski spacetime. In the particular case of an exactly face-on view, we have
\be
\label{eq:k4vec_faceon}
\mathbf{k} = \mathbf{e_t} + \boldsymbol{e_\theta}, \quad \mathrm{(face-on)}
\ee
such that the spatial component of the 4-vector points towards the negative Z axis, that is, towards the face-on observer.

Our final goal is to compute the emission EVPA,
so we do not need this null 4-vector, but rather its spacelike projection orthogonal
to the 4-velocity of the emitter, that is, in the rest frame of the emitter. This reads
\be
\label{eq:Kvec}
\mathbf{K} = \mathbf{k} + \left( \mathbf{k} \cdot \mathbf{u} \right) \, \mathbf{u}.
\ee
This simple relation is very crucial and virtually contains all
the results presented below. Even for a face-on observer, the actual
direction of photon emission does not lie along the vertical direction,
contrary to what is illustrated in the non-relativistic Fig.~\ref{fig:QUbasic}.
It acquires a toroidal component by means of the projection written
above, stemming from the toroidal component of $\mathbf{u}$. This is
simply the standard special relativistic aberration effect.

We can express
\be 
\label{eq:om} 
\mathbf{k} \cdot \mathbf{u} = -A \left(1 + \frac{\sin \iota \cos \pp}{\sqrt{r_0}} \right) \equiv -\omega,
\ee
where it is easy to check that $\omega$ coincides with
the norm of $\mathbf{K}$, that is, with the pulsation of the photon
as measured by the emitter.

\subsection{Vertical magnetic field}

Let us now restrict the discussion to an ambient vertical magnetic field. We want to derive an
analytic expression of the evolution of the emission EVPA with the orbital phase $\pp$.
For simplicity, we will consider here a pointlike hot spot
in the equatorial plane (so $\theta = \pi/2$ in all this section).
The unit vector along the magnetic
field direction reads
\be
\mathbf{\bar{B}} = -\boldsymbol{e_\theta}. 
\ee

Our goal is to express the emission EVPA, from Eq.~\ref{eq:chi}. Let us start by writing
\begin{align}
\label{eq:basis_vec}
  \mathbf{e_w} &=  \mathbf{e_X} = \cos \pp \, \boldsymbol{e_r} - \sin \pp \,\boldsymbol{e_\pp}, \\ \nn
  \boldsymbol{e_\delta} &= - \cos \iota \,\mathbf{e_Y} + \sin \iota \, \mathbf{e_Z} \\ \nn
  &= -\cos \iota \sin \pp \, \mathbf{e_r} - \sin \iota \, \boldsymbol{e_\theta} - \cos \iota \cos \pp \, \boldsymbol{e_\pp},  \\ \nn
\end{align}
where we note that we are working in the flat Minkowski spacetime, 
so the observer polarization basis is simply conserved along
the geodesic. We now need only the expression of the projection
of the magnetic field orthogonal to the direction of emission
\be
\mathbf{B_\perp} = \mathbf{\bar{B}} - \left( \mathbf{\bar{B}} \cdot \mathbf{\bar{K}}\right) \, \mathbf{\bar{K}},
\ee
where $\mathbf{\bar{K}} = \mathbf{K}/\omega$
is the unit vector along $\mathbf{K}$.

At this point, we have expressed all the quantities of interest
and can write the emission EVPA expression. The details of the computation
are not particularly illuminating, so we
provide them in Appendix~\ref{app:Minko}.
The final expression for the emission EVPA reads
\be
\label{eq:EVPAana}
%\mathrm{EVPA}
\chi_\mathrm{e}(\pp) = \frac{\pi}{2} - \mathrm{atan2} \left( \cos \iota \sin \pp \,\frac{A}{\omega \sqrt{r_0}}, \sin \iota + \cos^2 \iota \cos \pp \, \frac{A}{\omega \sqrt{r_0}} \right).
\ee

Let us first check what happens for an exactly face-on observer, $\iota = 0$.
In this case the expression simplifies considerably to $\chi_\mathrm{e}(\pp) = \pi/2 - \pp$.
It is clear from this expression that, as the hot spot rotates
with $\pp$ varying on a $2\pi$ interval, so will the emission EVPA. The emission EVPA will thus cover two times its domain of definition. And so will the observed EVPA, because the two quantities are nearly equal for our setup (see above).
So this will lead to a double QU loop seen by the distant observer.

\begin{figure}[htbp] 
\centering 
\includegraphics[width=0.5\textwidth]{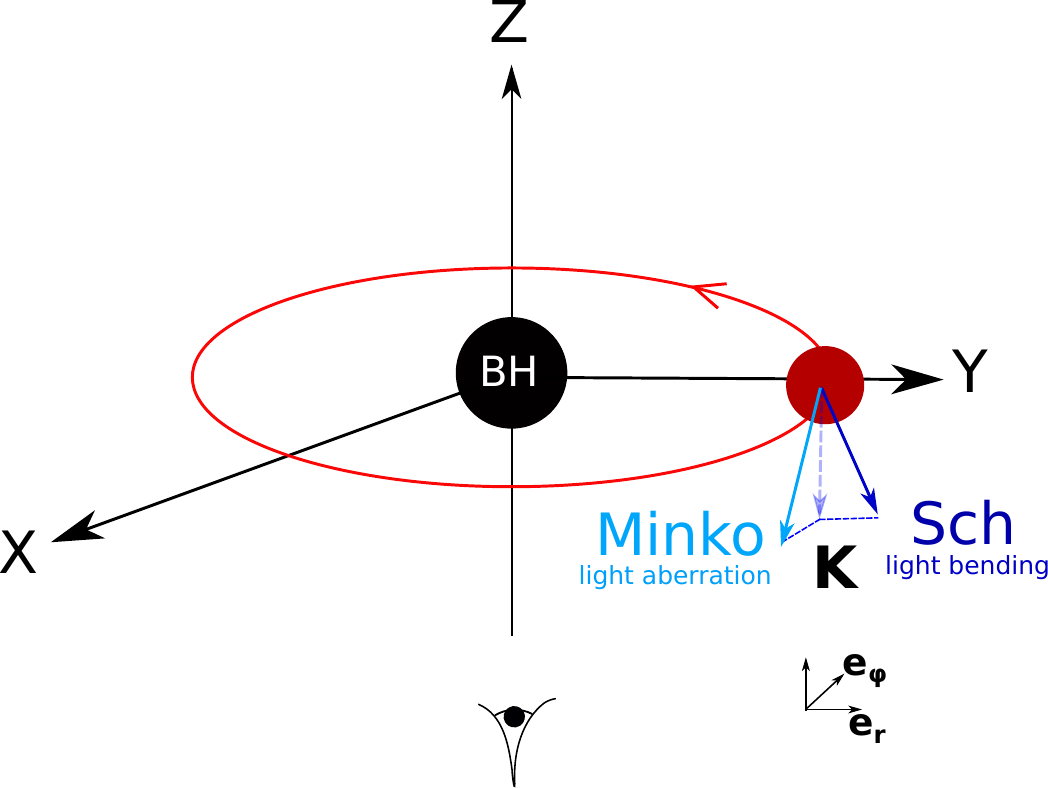} 
\caption{Effect of the spacetime geometry on the emission direction
$\mathbf{K}$. A hot spot (red disk) is orbiting around a black hole (black disk). The observer is located face-on towards the negative $Z$ axis. In a Newtonian spacetime, the direction of emission (i.e. the unit vector $\mathbf{K}$ along the projection of the null 4-vector $\mathbf{k}$
normal to the emitter's 4-velocity) is exactly vertical towards the
negative $Z$ axis (dashed pale blue arrow). This is the case illustrated
in the non-relativistic figure~\ref{fig:QUbasic}. 
Special relativistic light aberration
leads to an additional azimuthal component (solid light blue arrow). General relativistic
light bending leads to an additional radial component (solid dark blue arrow).
Note that the direction of emission in the Schwarzschild spacetime is along
the sum of the two solid arrows, given that the special relativistic aberration
is of course also included in the Schwarzschild geometry. The various vectors
are approximately to scale for a Keplerian hot spot at a few gravitational radii: the aberration and light bending effects are not
small corrections to an approximately vertical direction, they lead to
strong distorsions of the apparent emission direction (of order tens of percents).
} 
\label{fig:Kvec}
\end{figure}

This is the first non-intuitive conclusion of
our analysis: already in Minkowski, a face-on observer considering
a hot spot immersed in a vertical magnetic field will detect
a double QU loop signal. Note that the crucial difference between
the analysis developed in this section and the non-relativistic
analysis of section~\ref{sec:Newton} is the aberration affecting
the apparent direction of light propagation. The vector
$\mathbf{K}$ is not purely vertical, as is represented in
Fig.~\ref{fig:QUbasic}, it acquires a component in the equatorial
plane when projecting orthogonal to the relativistic 4-velocity
$\mathbf{u}$ of the emitter.
Figure~\ref{fig:Kvec} illustrates this.

We now turn to the discussion of a few important properties of Minkowski
QU loops in a vertical magnetic field, before discussing simulation
results.

\subsubsection{Emission EVPA symmetry}

Our emission EVPA expression has the following property
\be
\label{eq:2pi}
\chi_\mathrm{e}(\pp) = \pi - \chi_\mathrm{e}(2\pi - \pp) = - \chi_\mathrm{e}(2\pi - \pp)
\ee
where the second equality comes from the fact that the EVPA is
defined modulo $\pi$.
This relation means that the first half of the orbit $\pp \in [0,\pi]$
and the second half $\pp \in [\pi, 2 \pi]$ have the same EVPA evolution,
up to a sign difference. Equivalently, the EVPA orbital evolution is symmetric with respect to $\pp=\pi$, up to a sign. 

\subsubsection{QU loop mirror symmetry}
\label{sec:QUsym}

EVPA is not the only quantity that shows a symmetry in the orbital
evolution of the hot spot. The same goes for the photon's emitted energy
$\omega$. It is indeed obvious from Eq.~\ref{eq:om} that
\be
\label{eq:omega_sym}
\omega(2\pi - \pp) = \omega(\pp).
\ee
The same also goes for the angle 
\be
\theta_B = \mathrm{acos}\left( \mathbf{\bar{K}} \cdot \mathbf{\bar{B}} \right) 
\ee
between the magnetic field
and the photon's direction of emission. Indeed, Appendix~\ref{app:Minko}
shows that, for a vertical magnetic field,
\be
\label{eq:KB_vert}
\mathbf{\bar{K}} \cdot \mathbf{\bar{B}} = - \frac{\cos \iota}{\omega(\pp)}, %\quad \mathrm{[Vertical]}%\mathbf{K} \cdot \mathbf{B} &= - \frac{A}{\omega(\pp)} \left( \frac{1}{\sqrt{r_0} + \sin \iota \cos \pp}\right), \quad \mathrm{[Toroidal]} \\ \nn
\ee
where the $\pp$ dependence is made explicit. We thus have\be
\label{eq:thetaB_sym}
\theta_B(2\pi - \pp) = \theta_B(\pp).
\ee

The emitted flux only depends on the photon's emitted energy as well
as on the direction of emission relative to the magnetic field direction.
Indeed, for our circular orbit, all other physical quantities (density, magnetic
field magnitude, temperature) are constant. As a consequence,
Eqs~\ref{eq:omega_sym} and~\ref{eq:thetaB_sym} mean that the emitted
linearly polarized flux satisfies
\be
\label{eq:FLP_sym}
F_\mathrm{LP}(2\pi - \pp) = F_\mathrm{LP}(\pp).
\ee
Together with Eq.~\ref{eq:2pi}, and keeping in mind that for our setup the emission and observed EVPA are nearly equal, this relation leads to the conclusion
that the QU track in the Minkowski spacetime is symmetric with
respect to the horizontal axis.
Indeed, Fig.~\ref{fig:QUbasic} shows
that the linearly polarized flux and the double of the observed EVPA (compare to Eq.~\ref{eq:EVPAdef}) are the polar coordinates of the QU track.
For the rest of this article we will refer to this symmetry with respect to the horizontal Q axis as the QU loop mirror symmetry.

\subsubsection{Number of loops}

\begin{figure*}[htbp] 
\centering  
\includegraphics[width=\textwidth]{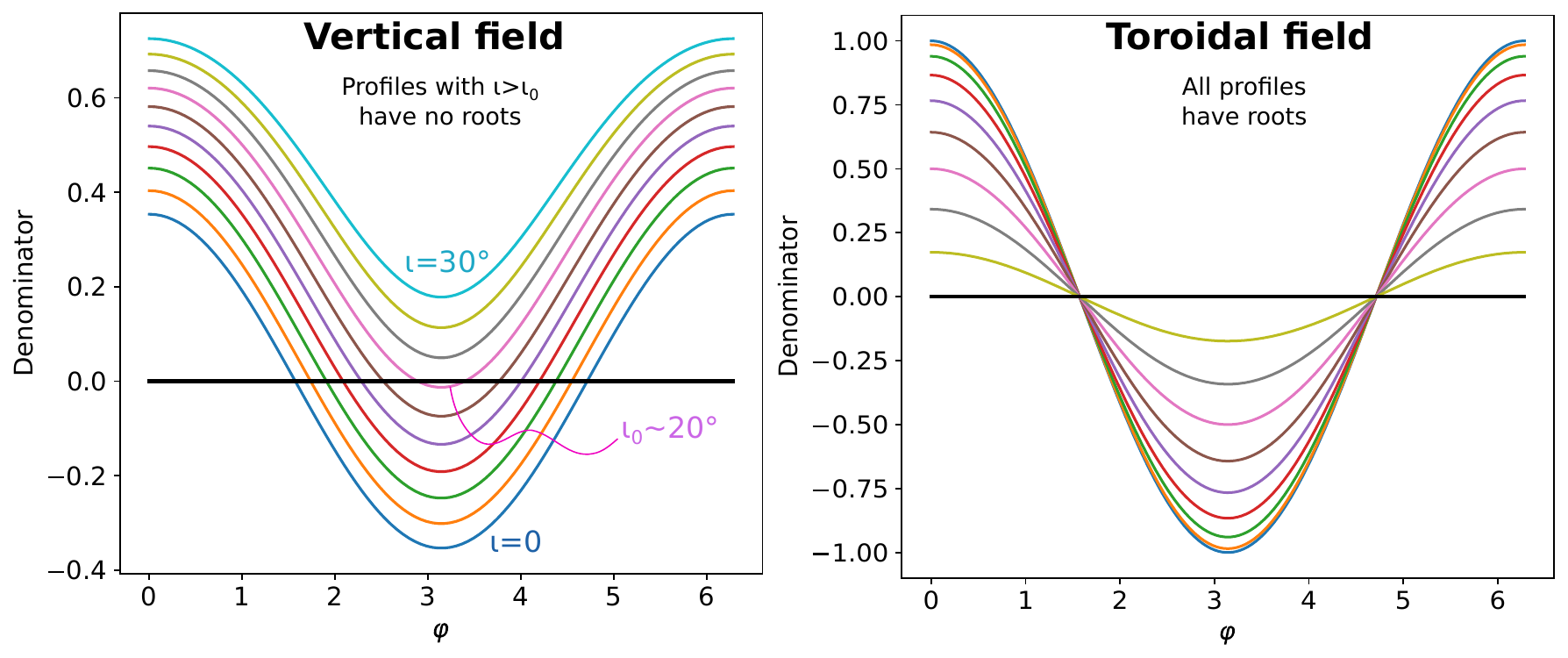} 
\caption{Denominator of the expression on the rhs of Eq.~\ref{eq:tanphi_vert} 
  (left panel, vertical field)
  and~\ref{eq:tanphi_tor} (right panel, toroidal field).
  These expressions are strongly dependent on the orbital radius,
  which is set to $r_0=8M$ here.
  The various colors encode various values of $\iota$ in 
  $[0,30^\circ]$ (left panel) or $[0,90^\circ]$ (right panel).
  In the vertical case, the denominator has a root only for
  $\iota < \iota_0(r_0)$, and this critical angle verifies 
  $\iota_0\approx 20^\circ$ for $r_0=8M$.
  The condition $\iota < \iota_0(r_0)$
  will lead to two QU loops, while higher inclinations will
  lead to a single QU loop. In the toroidal case, all values of $\iota$
  lead to the existence of two roots, so there will always be two loops,
  whatever the inclination.}  
\label{fig:root}
\end{figure*}

\begin{figure}[htbp] 
\centering  
\includegraphics[width=0.5\textwidth]{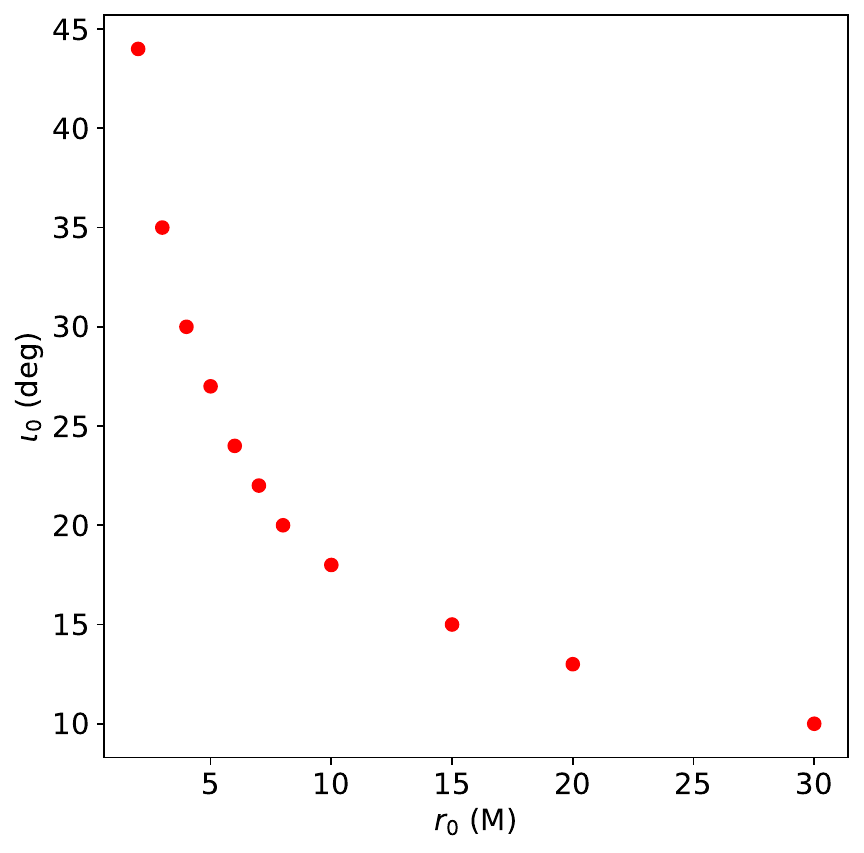} 
\caption{Evolution of the limit inclination angle $\iota_0$ (see Eq.~\ref{eq:iota0}), which separates double ($\iota<\iota_0$) and single ($\iota>\iota_0$) QU loops in Minkowski spacetime for a vertical magnetic field, with the hot spot orbital radius $r_0$. This angle converges towards $0$ as $r_0$ increases.}
\label{fig:i0r0}
\end{figure}

The emission EVPA orbital evolution $\chi_\mathrm{e}(\pp)$ is dictated by Eq.~\ref{eq:EVPAana},
and is symmetric with respect to $\pp=\pi$ up to a sign.
Thus, if the full allowed range of EVPA,
$[-\pi/2 , \pi/2]$, is covered in the first half of the orbit,
then it will be covered again in the second half, leading to
two QU loops. This can happen provided that the EVPA visits
all possible values in $[-\pi/2 , \pi/2]$ during the first orbit,
so if its tangent reaches infinity.
There will thus be two QU loops provided that
\be
\label{eq:tanphi_vert}
\frac{\mathbf{B_\perp} \cdot \mathbf{e_w} }{\mathbf{B_\perp} \cdot \boldsymbol{e_\delta}} = \frac{\cos \iota \sin \pp \,\frac{A}{\omega \sqrt{r_0}}}{\sin \iota + \cos^2 \iota \cos \pp \, \frac{A}{\omega \sqrt{r_0}}}
\ee
varies between $-\infty$ and $+ \infty$ when $\pp$ varies between
$0$ and $\pi$. This quantity will reach infinity provided that the denominator
\be
\sin \iota + \cos^2 \iota \cos \pp \, \frac{A}{\omega \sqrt{r_0}} = 0,
\ee
considered as an equation for the variable $\pp$ with a given
inclination $\iota$, has a root for some value of $\pp$.
Note that this is not such a trivial equation as it
might seem, because $\omega$ depends on $\iota$, see Eq.~\ref{eq:om}.
%When $\iota=0$ (face-on observer), there is an obvious root for $\pp=\pi/2$.
By examining this function numerically it is easy
to show that it has a root only when
\be
\label{eq:iota0}
\iota < \iota_0(r_0)
\ee
which is the condition for obtaining two loops in a vertical
magnetic field, in Minkowski spacetime. The limiting angle
$\iota_0$ depends on the orbital radius $r_0$,
the dependence being illustrated in Fig.~\ref{fig:i0r0}.
This is illustrated
in the left panel of Fig.~\ref{fig:root}.
We note that the existence of such a limit angle behavior for
the existence of one or two loops has already been discussed
in the Schwarzschild context by~\citet{gelles21}, see their Fig.~9,
but without the analytical treatment that we provide here building on
the simplicity of the Minkowski geometry.

%\textbf{TBD: discuss the evolution of $\iota_0$ with $r_0$. It goes
%  to zero when $r_0$ increases (not  because we go away from the strong
%  gravitational field region, but because the velocity decreases and becomes
%  less relativistic, so less aberration effect, converges to Newtonian).}

This is the second conclusion of this section: QU loops of the
Minkowski spacetime in a vertical magnetic field
share the exact same property as already discussed
in the Kerr context by several authors~\citep{jimenez20,gelles21,vos22},
that is, the existence of either one or two loops depending on the
inclination and on the orbital radius. To our knowledge, the relation between this behavior and
the special relativistic aberration effect has not been discussed in the literature to date.
%The simplicity of the Minkowski geometry allows us to
%provide a complete analytical understanding of this behavior.
%We conclude that a hot spot immersed in a vertical magnetic field
%in Minkowski spacetime
%can lead to either one or two QU loops, depending on the inclination,
%lower inclination leading to two loops, higher inclination to one loop.

%When two loops are drawn in the QU plane, they typically have different
%sizes, with one outer and one inner loop. This is due to the variation
%of polarized flux, that follows the evolution of the total flux.
%The polar radius in the QU plane being equal to the linearly
%polarized flux, the loop size follows the flux variation, and we
%discuss below the rather intricate origin of this flux variation.

%\textbf{It seems from the simu that the QU loops of Minkowski
%  is always symetric wrt the horizontal, not yet clear why, would
%  be nice to demonstrate.} \mw{must be gravitational lensing!}
%  \fv{No it is a consequence of Eq.~\ref{eq:2pi}, see the text edition after this equation.}

\subsubsection{Simulated QU loops}

\begin{figure*}[htbp] 
\centering  
\includegraphics[width=0.8\textwidth]{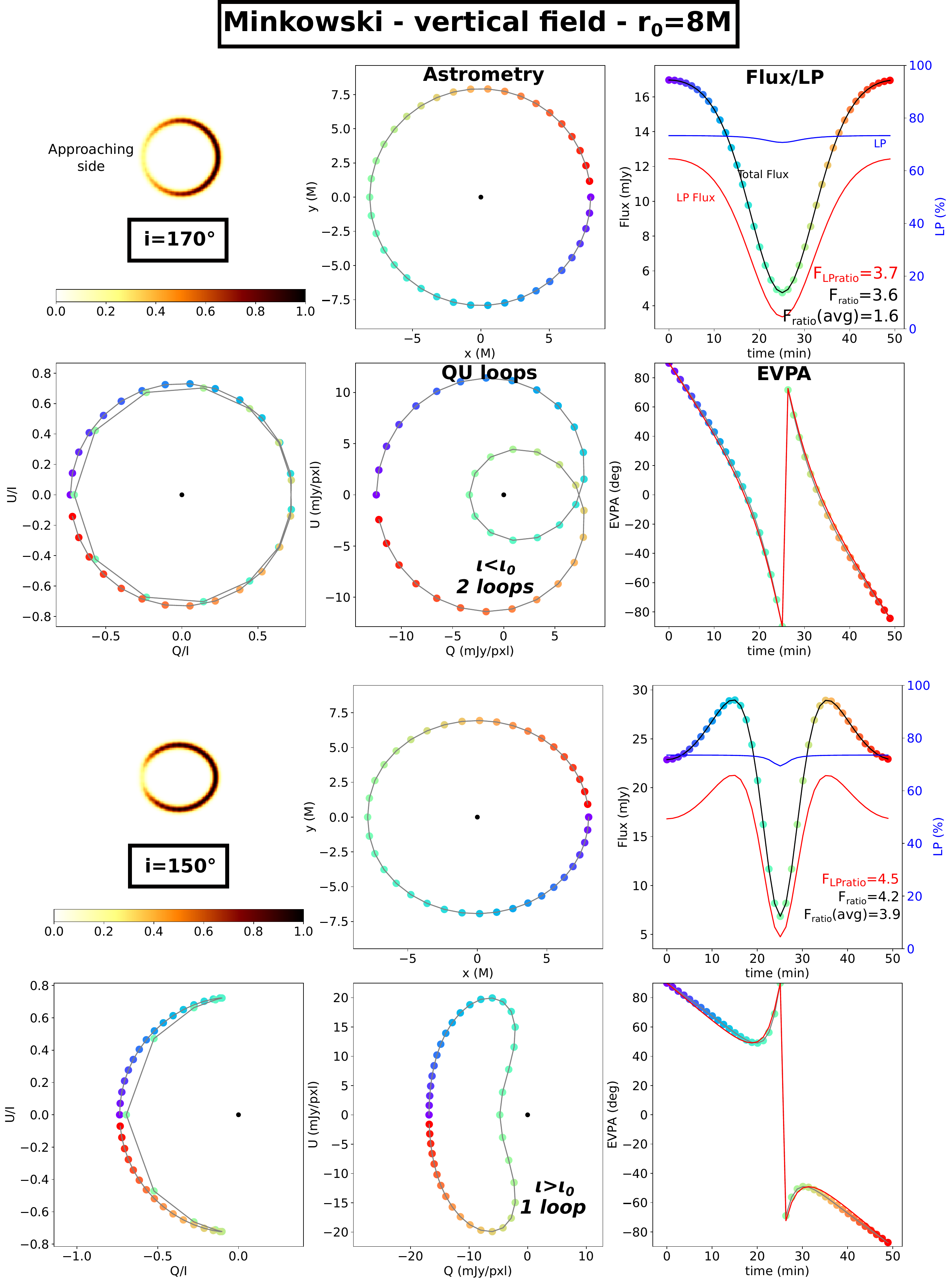} 
\caption{Minkowski QU loops in vertical magnetic field.
  The top six panels are computed for $\iota = 10^\circ < \iota_0$,
  where $\iota_0$ is defined in Eq.~\ref{eq:iota0} and defines the highest
  angle for which there should be two QU loops. The bottom six panels
  are computed for $\iota = 10^\circ > \iota_0$. The six panels represent
  the following quantities. 
  Top-left: the summed images of the hot spot in normalized intensity;
  we note that the color coding is inverted to improve
  the readability, darker color means more intense emission.
  Top-middle:
  the astrometric track on sky; in this panel and the next ones, color
  codes for time, from violet to red, clockwise motion on sky. 
  Top-right:
  the total flux (colored dots), linearly polarized flux
  ($F_\mathrm{LP} = \sqrt{Q^2+U^2}$, red curve),
  and linear polarization fraction (LP, in percent, blue curve)
  evolution; the flux
  ratio $F_\mathrm{ratio}$
  (maximum over minimum fluxes) is provided in the bottom-right
  corner of the panel, together with the linearly polarized flux
  ratio (written in red), and the flux ratio
  $F_\mathrm{ratio}\mathrm{(avg)}$ obtained 
  after averaging over the angular dependence of the radiative transfer
  coefficients (the $\sin \theta_B$ dependence).
  We note that the density and temperature of the hot spot have been
  chosen such that the low-inclination, vertical magnetic field
  near infrared flux peaks around $10$~mJy.
  Bottom-left: the (Q/I,U/I)
  plane. Bottom-middle: the (Q,U) plane, to which we refer when
  discussing the QU loops.
  Bottom-right: the observed EVPA
  evolution; the red profile shows
  the emission EVPA evolution as predicted by the analytic model
  derived in Eq.~\ref{eq:EVPAana}.
  As predicted, the upper case shows two QU loops,
  while the bottom one shows only one loop. 
}
\label{fig:QUvert_Minko}
\end{figure*}

Figure~\ref{fig:QUvert_Minko} illustrates these findings by showing the
results of a polarized ray-tracing calculation in Minkowski spacetime
for a hot spot seen under an inclination smaller and bigger than
the critical angle $\iota_0 \approx 20^\circ$ for $r_0=8M$. As predicted,
we obtain respectively two and one QU loops in these cases. The EVPA
evolution follows very precisely the analytical prediction of
Eq.~\ref{eq:EVPAana} at low inclination, which validates our calculation, and is at the
same time a non-trivial consistency test of our polarized radiative
transfer. We note that this is a clear demonstration of the near equality between the emitted and observed EVPA, because the colored dots and the red profile of the EVPA panel in Fig.~\ref{fig:QUvert_Minko} respectively represent an observed and emitted EVPA.
It is also interesting to note that, although the analytical and numerical EVPA profiles remain similar, they are clearly more different at higher inclination, $\iota=30^\circ$. This is not due to a limitation of the precision of the numerical integration. Instead, the differences are related to the Roemer effect, due to the finite velocity of light, that is not taken into account in the analytical profile. As a consequence, the numerical data lead the analytical profile in the first half orbit (where the hot spot is further away from the observer), while it lags behind the analytical profile in the second half orbit (where the hot spot is closer to the observer). As expected, the exact same behavior occurs for a toroidal magnetic field.
Moreover, the QU track is mirror-symmetric, as predicted above. %\mw{[does Roemer effect break the mirror symmetry then? "up to the Roemer effect"?]} \fv{No I don't think it breaks it but it's too late to give a formal demonstration.} \mw{but if analytic are symmetric, and numerical are a bit different because light speed, then doesn't it mean the numerical/"real" are not perfectly symmetric?} \fv{If you check carefully the curve, the numerical data, although different from the analytics, is still symmetric. The shift is mirrored in the first-half and second-half orbits.} \mw{Nice, interesting that lagging in one half would look the same/ symmetric as advancing in the other half. Probably can be understood if one thinks about this a bit. I'm fine with the text now.}

\begin{figure*}[htbp] 
\centering  
\includegraphics[width=\textwidth]{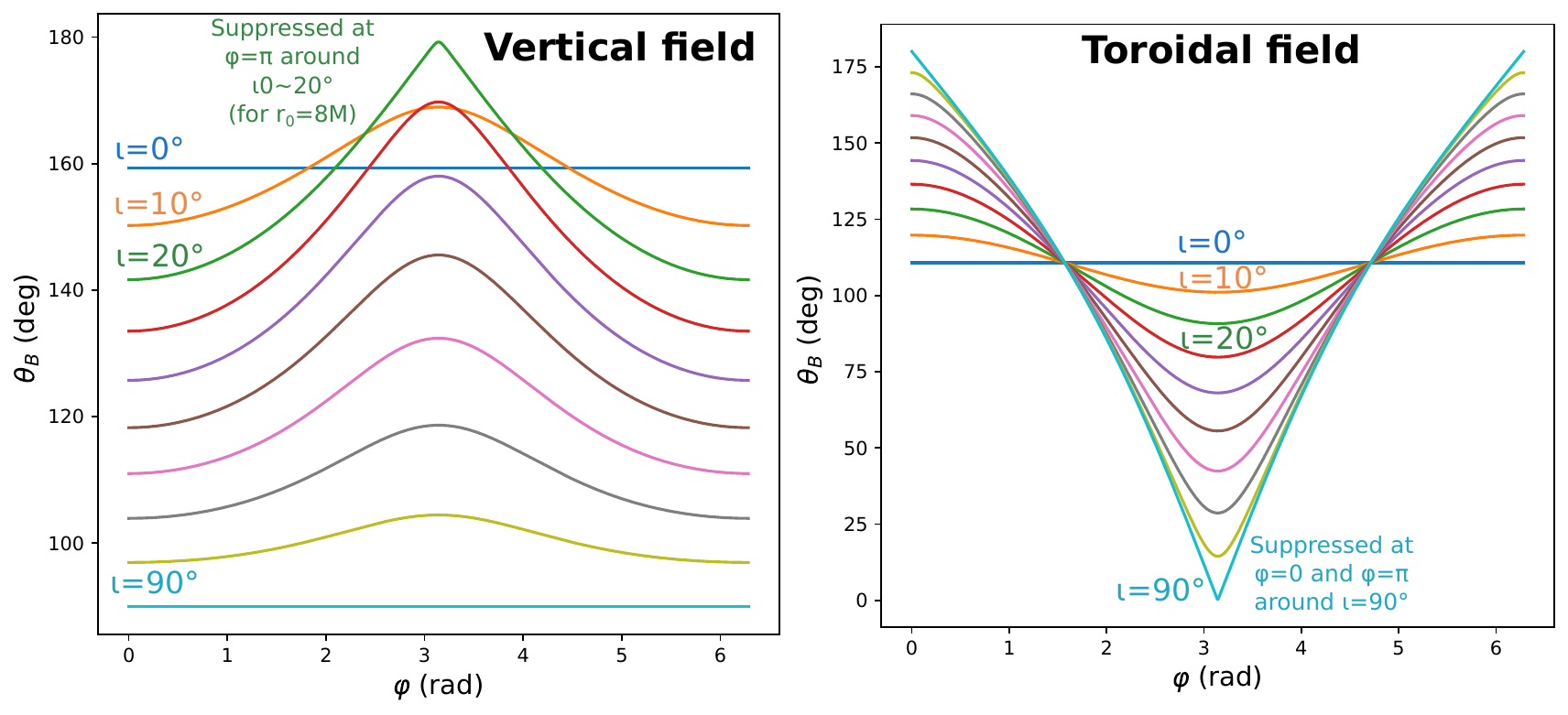} 
\caption{Minkowski evolution of the angle $\theta_B$, the angle between 
  the magnetic field and the emission direction in the emitter's frame, for $r_0=8M$.
  The various colors encode various inclinations
  $\iota$ between $0$ (face-on, dark blue) and
  $90^\circ$ (edge-on, light blue), with a $10^\circ$ step.
  The synchrotron emission is suppressed
  at $\theta_B=0 [\pi]$, so we conclude that the orbital phase
  $\pp = \pi$ (corresponding to the left part of the image,
  towards the East direction) is strongly suppressed around
  $\iota = \iota_0 \approx 20^\circ$. We remind that this angle 
  $\iota_0$ depends on the orbital radius, here $r_0=8M$.
}
\label{fig:thetaB}
\end{figure*}

It is interesting to note that the evolution of the observed flux
might seem counter-intuitive. Indeed, the source is approaching the
observer on the left part of the trajectory (East side). But the
flux evolution (upper-right panel of Fig.~\ref{fig:QUvert_Minko})
shows that contrary to what relativistic beaming intuition would 
suggest, the flux is actually at minimum on the approaching side.
This is a consequence of the $\sin \theta_B$ dependence of the
synchrotron radiative transfer coefficients, see Eq.~\ref{eq:jnuPolar}.
%where $\theta_B$ is the angle between the direction 
%of the magnetic field and the direction of emission in the emitter's frame.
This angle is close to $0 \,[\pi]$ on the left side of the sky plane
(which corresponds to an orbital phase $\pp=\pi$),
as demonstrated by the analytical profiles of the left
panel of Fig.~\ref{fig:thetaB}.
These profiles represent the orbital phase evolution of 
$\theta_B = \mathrm{acos} \left( \mathbf{\bar{K} \cdot \mathbf{\bar{B}}}\right)$,
the expression of which is known analytically from the formulas provided in
Appendix~\ref{app:Minko}.
We note that around $\iota=\iota_0 \approx 20^\circ$ (for $r_0=8M$),
the influence of the
$\theta_B$ dependence of the emission
not only mitigates the relativistic beaming, but inverses
the tendency by leading to a light curve that peaks on the receding side.
We have checked that if one averages over $\sin \theta_B$ (that is, if one considers an isotropized emission), the usual
flux profile, peaking on the approaching side, is recovered. The Doppler effect cannot be responsible for this strong flux depletion at the orbital phase $\pp=\pi$, because the emitted frequency is at minimum at the orbital phase $\pp=\pi$ (see the top-left panel of Fig.~\ref{fig:Qty_QUsym}), so the emitted Doppler-shifted flux is actually maximised there (see Eq.~\ref{eq:jnuPolar}).
%We note that the flux ratio between the maximum and minimum observed
%value is equal to $\approx 4$ in the case illustrated in the upper-right
%panel of Fig.~\ref{fig:QUvert_Minko}), while it drops to only $\approx 1.6$
%when averaging over $\sin \theta_B$. This shows that, for this
%low-inclination, vertical magnetic field case, the angular dependence
%of the radiative transfer coefficients have a much stronger impact on the
%flux variation than the relativistic beaming, and that rather strong
%flux excursions can be observed even at rather low inclination.
Figure~\ref{fig:thetaB} shows that this behavior is specific to the
low inclination. Higher inclination progressively leads to the
{more intuitive} situation dominated by relativistic beaming. This is very natural: a vertical magnetic field seen at low inclination leads
to $\theta_B$ angles around $0 [\pi]$, where the $\sin \theta_B$
dependence of the radiative transfer coefficient has a strong impact, while at high inclination,
$\theta_B$ varies around $\pi/2$, where this dependence is weaker.
This is the third conclusion of this section: in the Minkowski spacetime and
for a vertical magnetic field, the flux variation is driven by
the angular dependence of the synchrotron radiative transfer
coefficients at low inclination,
%(that can lead to a strong flux
%contrast of $\approx 4$, contrary to what intuition might suggest
%at low inclination), 
and by relativistic beaming
at high inclination.

The linear polarization of our hot spot is always
very high, of order $75\%$, which is twice as high as the typically
observed near infrared values~\citep[e.g.][]{gravity18}. This is due to the very simple setup
that we consider, with a small isolated emitting body. A more realistic
scenario~\citep[see e.g. App.~B of][]{gravity23}, with a more extended or distorted structure, and
the addition of larger-scale quiescent emission, would recover a more realistic level of linear polarization.

\subsection{Toroidal magnetic field}

The exact same computation that we presented in the last section for a vertical 
magnetic field can be performed for a toroidal magnetic field. Starting
from Eq.~\ref{eq:Bdef}, and specializing to the Minkowski spacetime  
in the equatorial plane, we obtain
\be
\mathbf{\bar{B}} = A \left(\frac{\mathbf{e_t}}{\sqrt{r_0}} + \boldsymbol{e_\pp} \right),
\ee
which is a unit spacelike vector normal to $\mathbf{u}$.

We refer the reader to Appendix~\ref{app:Minko} for the details of the
computation and simply give here the final result
\be
\label{eq:EVPA_tor}
  \chi_\mathrm{e}(\pp) = \frac{\pi}{2} - \mathrm{atan2} \left(\sin \pp \left[ C \, \frac{A \omega}{\sqrt{r_0}} - 1 \right], \cos \pp \cos \iota \left[ C \, \frac{A \omega}{\sqrt{r_0}} - 1 \right]\right),
\ee
where $C = 1/\omega^2 \left(1/\sqrt{r_0} + \sin \iota \cos \pp \right)$.

Similar properties as in the vertical case can be derived in the exact
same way as presented in the previous section. In particular, the relation
\be
\label{eq:2pibis}
\chi_\mathrm{e}(\pp) = - \chi_\mathrm{e}(2\pi - \pp)
\ee
still holds, and the QU loop mirror symmetry as well, which is due
to the symmetry of the expression of the emission angle
for a toroidal magnetic field, derived in App.~\ref{app:Minko},
\be
\label{eq:KB_tor}
\mathbf{\bar{K}} \cdot \mathbf{\bar{B}} = - \frac{A}{\omega(\pp)} \left( \frac{1}{\sqrt{r_0} + \sin \iota \cos \pp}\right), %\quad \mathrm{[Toroidal]} \\ \nn
\ee
leading to the same property as in Eq.~\ref{eq:thetaB_sym}.

The number of QU loops can be studied following
the same reasoning as in the previous section. This leads to studying the range
of variation of the simple expression
\be
\label{eq:tanphi_tor}
\frac{\mathbf{B_\perp} \cdot \mathbf{e_w} }{\mathbf{B_\perp} \cdot \boldsymbol{e_\delta}} = \frac{\sin \pp }{\cos \pp \, \cos \iota},
\ee
and in particular the roots of the denominator
\be
\cos \pp \cos \iota = 0
\ee
as $\pp$ varies in $[0,2\pi]$.
Obviously here, there are always two roots at $\pp=\pi/2, 3\pi/2$ whatever the inclination (see the illustration in the right panel of Fig.~\ref{fig:root}), leading
to the existence of two QU loops for all inclinations.
\begin{figure*}[htbp] 
\centering  
\includegraphics[width=0.8\textwidth]{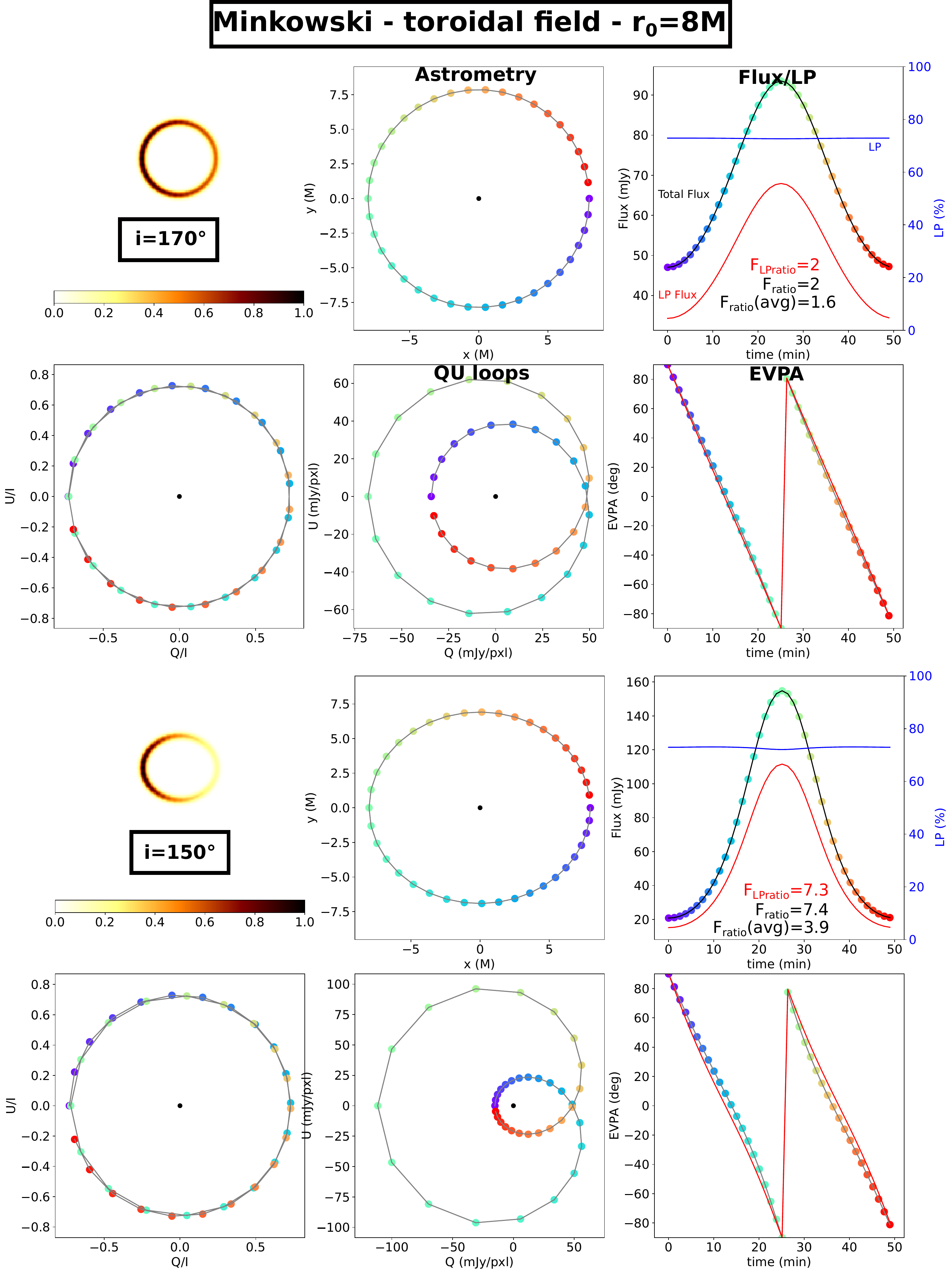} 
\caption{Same as Fig.~\ref{fig:QUvert_Minko} for a toroidal magnetic field.
  Two QU loops are present for both inclinations, contrary to the vertical
  case of Fig.~\ref{fig:QUvert_Minko}, in agreement with our analytical
  derivation.
}
\label{fig:QUtor_Minko}
\end{figure*}

Figure~\ref{fig:QUtor_Minko} shows simulations of Minkowski QU loops
in a toroidal magnetic field, and confirms
the existence of a double loop for the same two values of inclinations
that lead to either one or two loops in the vertical field case,
in perfect agreement with the results above.
The numerical profile of the EVPA exactly matches the analytics at low inclination, and is slightly offset with respect to the analytics at higher inclination, because of the Roemer effect, as discussed for the vertical case.
We note that contrary to the vertical case, the flux evolution
appears to follow here the standard relativistic beaming intuition,
with the flux peaking at the approaching side and the flux ratio
increasing with the inclination. This is because
the $\theta_B$ dependence is very weak at low inclination
$\iota \lesssim 45^\circ$, as demonstrated by the right panel
of Fig.~\ref{fig:thetaB}. At high inclination on the contrary,
the $\theta_B$ dependence becomes very strong and would counteract
the beaming effect. This dependence is very natural: at low inclination, a toroidal magnetic field leads to $\theta_B$ angles
much closer to $\pi/2$ than to $0 [\pi]$, while at high inclination,
the contrary is true. The dependence is reversed compared to the
vertical magnetic field case.

\section{QU loops in Schwarzschild spacetime}
\label{sec:Sch}

This section presents QU loops computations considering the same setups as illustrated in the previous section, but taking into account the spacetime curvature associated with the Schwarzschild geometry. We stress that we consider only the primary image and do not include the secondary or higher-order images formed by the extremely lensed photons executing at least half an orbit around the black hole \citep[e.g.,][]{Johnson2020}. These higher-order images do not change the main qualitative features of the QU loops but have an impact at a finer level, at the low and moderate inclinations that we consider here, see for instance~\citet{gelles21,wielgus22}. This simplification allows to reduce the needed imaging resolution.

\begin{figure*}[htbp] 
\centering  
\includegraphics[width=0.8\textwidth]{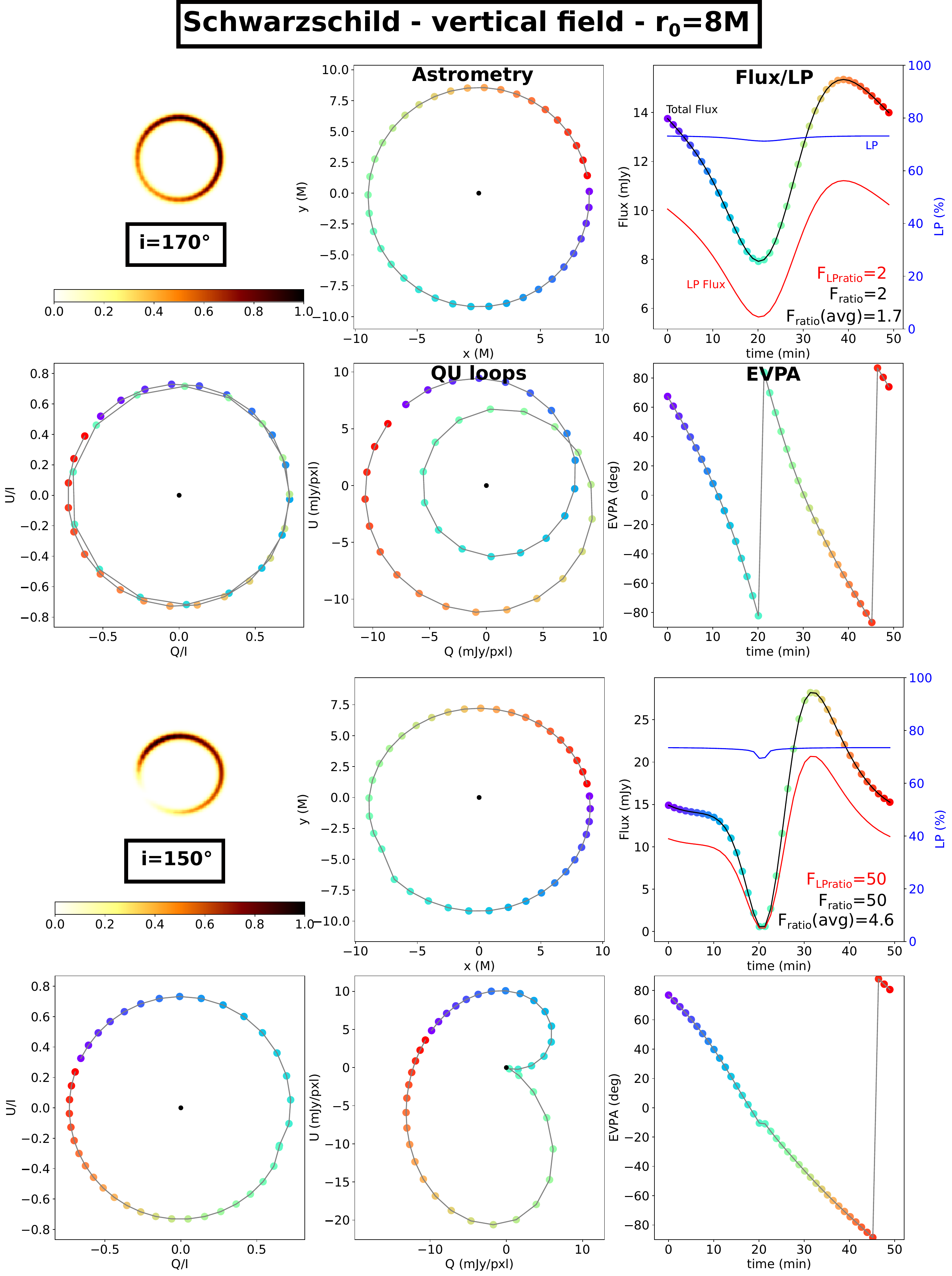} 
\caption{Same as Fig.~\ref{fig:QUvert_Minko} but in the Schwarzschild spacetime,
  for a vertical magnetic field. It might seem surprising that there is a
  small kick on the astrometric path towards the South-East. This is due to
  the dependence of the radiative transfer coefficients on
  $\sin \theta_B$, where $\theta_B$ is the angle between the direction
  of the magnetic field and the direction of emission, see~\citet{marszewski21}.
  The upper-left panel clearly shows a flux depletion towards the
  South-East, due to this effect. At this orbital phase, the direction of
  emission in the emitter's frame, $\mathbf{K}$, becomes vertical and
  parallel to the magnetic field. Due to the combination of special-relativistic
  aberration and general-relativistic lensing effects,
  the direction of $\mathbf{K}$ varies
  with orbital phase. The QU loops of this figure should be compared to that
  of Fig.~\ref{fig:QUvert_Minko}: the similarity is striking.
}
\label{fig:QUvert_Sch}
\end{figure*}

\begin{figure*}[htbp] 
\centering  
\includegraphics[width=0.8\textwidth]{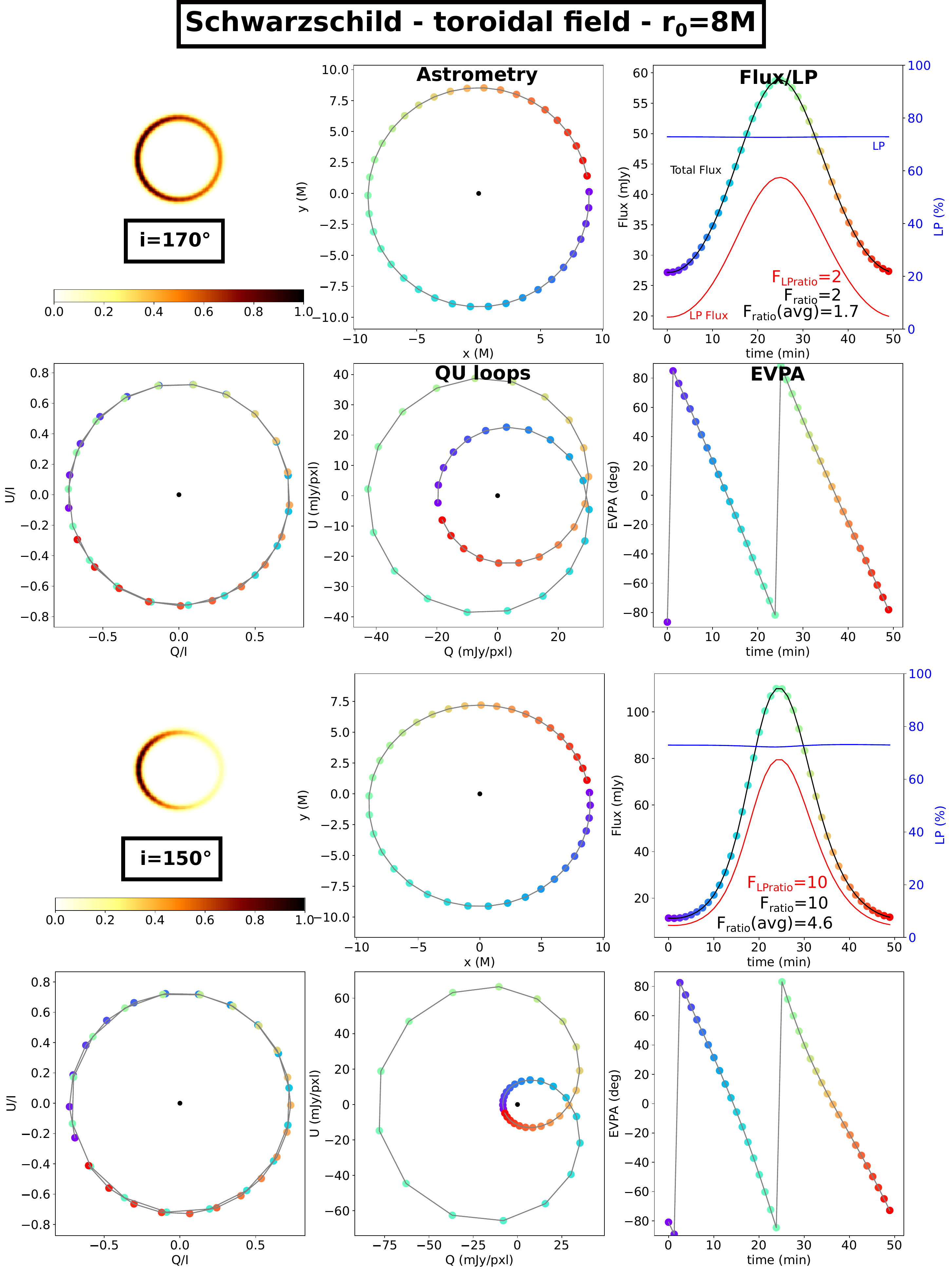} 
\caption{Same as Fig.~\ref{fig:QUvert_Minko} but in the Schwarzschild spacetime,
  for a toroidal magnetic field. The QU loops of this figure should be compared to that
  of Fig.~\ref{fig:QUtor_Minko}: the similarity is striking.
}
\label{fig:QUtor_Sch}
\end{figure*}

Figures~\ref{fig:QUvert_Sch}
and~\ref{fig:QUtor_Sch} show these QU loops, in the case of a
vertical or toroidal magnetic field, respectively. Interestingly, for most cases, there is no pronounced difference between these Schwarzschild QU loops and their Minkowski counterparts computed in the previous section. 
%it is probably not obvious to make
%an observational clear difference between these Schwarzschild QU loops
%and their Minkowski counterparts computed in the previous section.
The main features of the loops are already present in flat spacetime
and while the light bending changes detailed values of the observables,
it has little impact on the general picture. {Naturally, we would expect a more significant impact in case of an orbital radius smaller than the observationally motivated $r_0 = 8 r_g$ that we have assumed.  }

Regarding the flux variation, we note the same behavior of the
Schwarzschild/vertical cases as discussed for their Minkowski
counterparts. The flux goes through its minimum at the hot spot approaching side, due to the $\theta_B$ dependence of the
radiative transfer coefficients. We note 
that the Schwarzschild/vertical case,
seen at $\iota=10^\circ$, shows a smaller flux variation than its
Minkowski counterpart (factor of $\approx 2$ versus a factor
of $\approx 4$ peak-to-peak ratio). This is because the value of
$\theta_B$ never goes as close to $\pi$ in the Schwarzschild
case as in the Minkowski case. As a consequence, the flux minimum
is higher in the Schwarzschild case. On the contrary, for
$\iota=30^\circ$, the value of $\theta_B$ in the Schwarzschild
case goes through nearly exactly $\pi$, leading to a flux
minimum approaching zero, contrary to the Minkowski case that
keeps $\theta_B$ further from $\pi$. This explains the extreme
flux ratio (factor of $50$!) for the Schwarzschild/vertical
case at $\iota=30^\circ$. 
The Schwarzschild/toroidal case is also similar to the corresponding
Minkowski setup in the sense that the flux variation is dominated
by relativistic beaming with the usual flux maximum at the
approaching side of the orbit. The flux ratios are similar for
Minkowski and for Schwarzschild, showing that the special-relativistic beaming effect is the dominant flux-driving mechanism. 

\section{Comparing Schwarzschild and Minkowski QU loops}

\begin{figure*}[htbp]  
\centering  
\includegraphics[width=\textwidth]{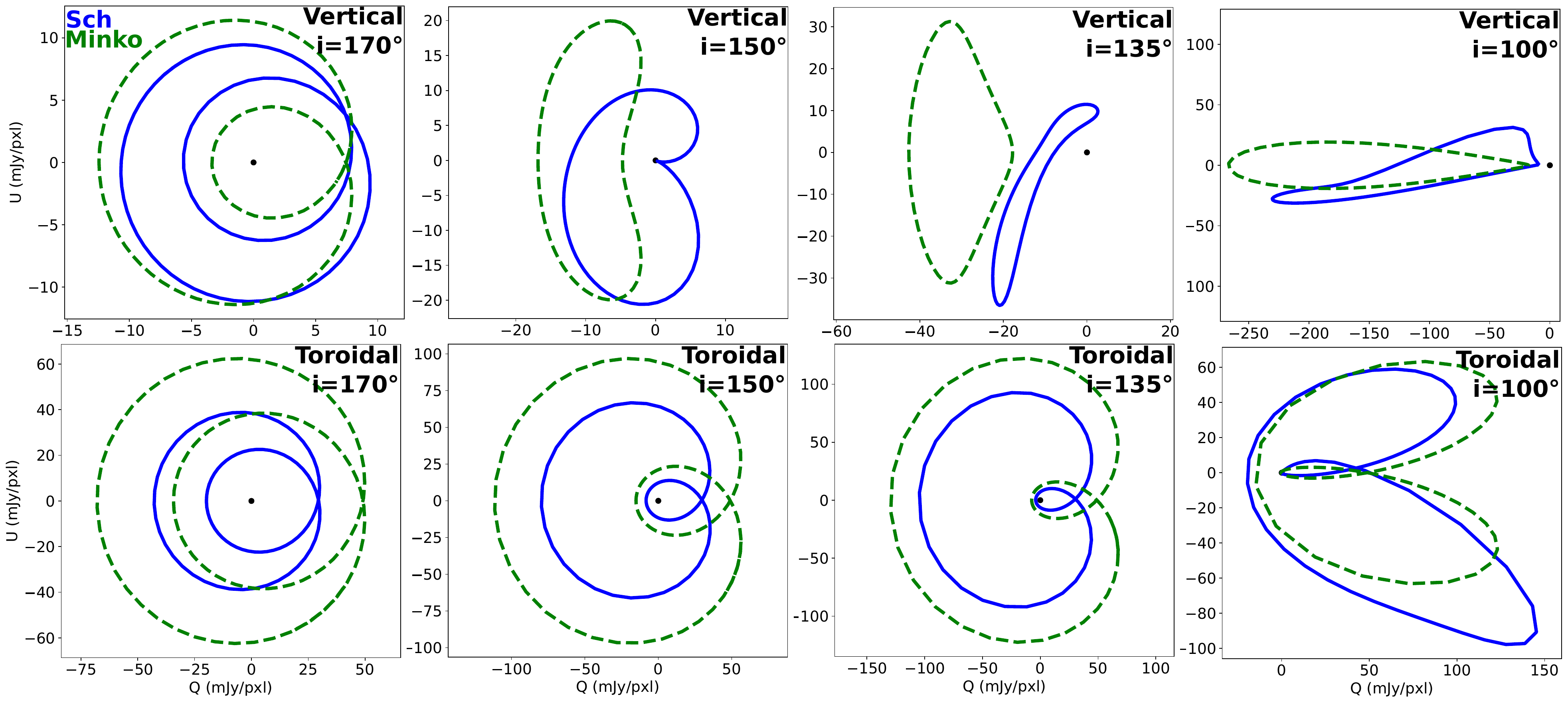} 
\caption{Comparison of the QU loops computed in the Schwarzschild (solid blue) and Minkowski (dashed green) spacetimes. The magnetic field is vertical for the top row and toroidal for the bottom row. The inclination increases from left to right and is specified in the top-right corner of each panel. Mind the
different scalings of the various panels. 
}
\label{fig:QUcompare}
\end{figure*}

Figure~\ref{fig:QUcompare} shows a comparison of the QU loops computed in the Schwarzschild and Minkowski spacetimes, that were presented in Figs.~\ref{fig:QUvert_Minko}-~\ref{fig:QUtor_Sch}, as well as two higher
inclination cases, $\iota=45^\circ, 80^\circ$. This figure again shows that 
flat-space and curved-space QU loops are very similar for most cases.
However, there is one important property that we demonstrated in the
Minkowski case (see Section~\ref{sec:QUsym}), the QU loop mirror symmetry, which is lost in Schwarzschild
as inclination increases. 
This is a direct consequence of light bending. We note that the QU loop fitted to the high-sensitivity ALMA observations appears strongly asymmetric \citep{wielgus22}.
%This is very obvious for the vertical case.
%This is less obvious for the toroidal case, but would happen at higher
%inclination ($\iota \gtrsim 80^\circ$). Schwarzschild QU loops are not symmetric in general,
%and this is a direct consequence of light bending. 

%Indeed, let us consider two positions along the hot spot trajectory, that are symmetric with respect to $\pp=\pi$ (see Fig.~\ref{fig:QUsym}).
%The wavevectors $\mathbf{k}$ along the null geodesic connecting the hot spot to the distant observer corresponding to these two positions are the same
%in the Minkowski spacetime. However they differ in Schwarzschild, due to
%light bending. This difference will translate in a difference in 
%the EVPA, the emitted energy $\omega$, and the emission angle $\theta_B$,
%such that the Schwarzschild spacetime does not share the same property
%of QU loops symmetry as in Minkowski. This is a specific Schwarzschild 
%feature linked to light bending, and might be used as a probe of
%spacetime curvature when fitting observed data. It is particularly interesting
%to note that this Schwarzschild QU loop asymmetry is much more pronounced
%for a vertical magnetic field, which is a configuration favored by
%recent observations~\citep{wielgus22}.
%\fv{To be discussed.}

\begin{figure*}[htbp]  
\centering  
\includegraphics[width=\textwidth]{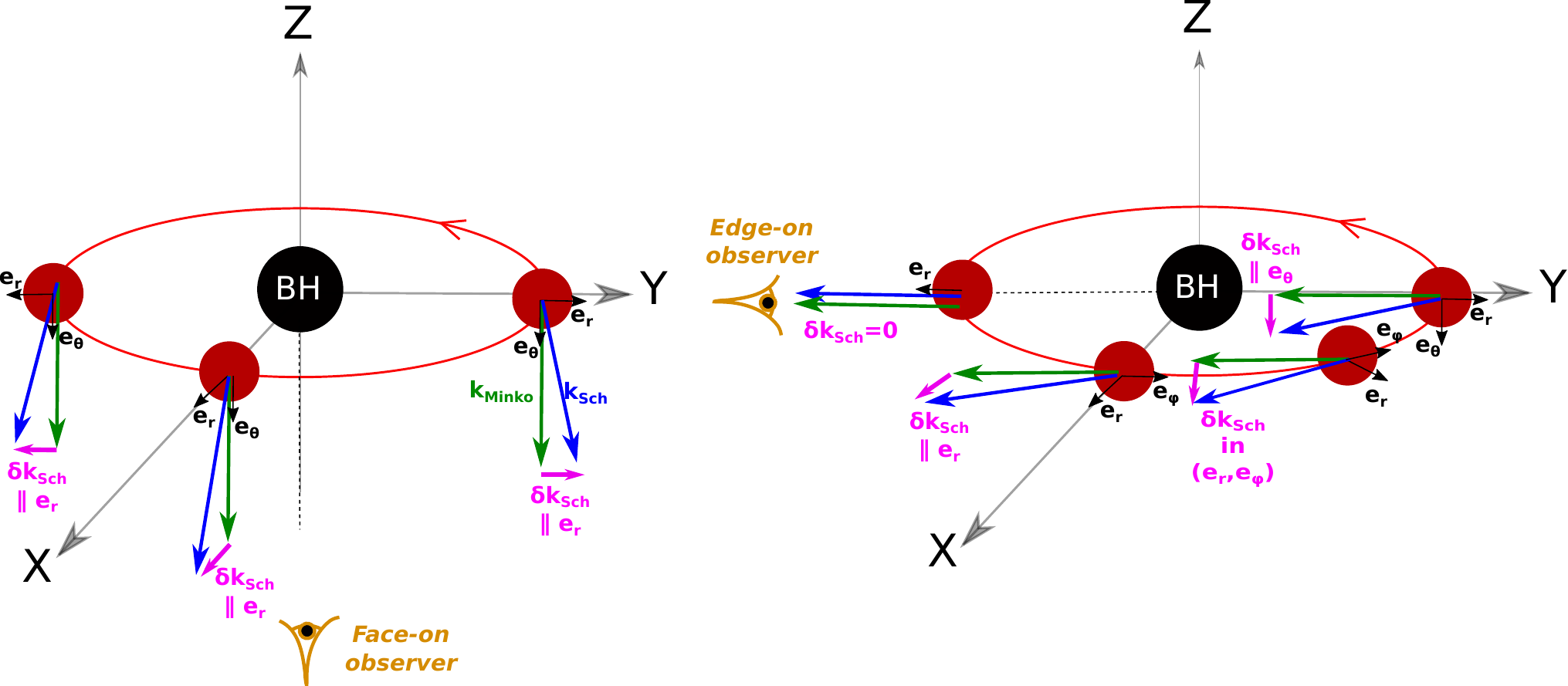} 
\caption{Lensing and asymmetry of Schwarzschild QU loops. The left panel is depicted at zero inclination, the right one is edge-on. The green arrows show the wavevectors $\mathbf{k}$ in Minkowski spacetime that connect the hot spot to the observer. The blue
arrows show the corresponding wavevectors for the Schwarzschild case.
They differ from Minkowski due to light bending, which adds a shift to the wavevector, depicted in pink. This shift vector is constant with orbital phase and along the positive radial direction at zero inclination. It varies a lot with orbital phase for edge-on view, from being zero at the closest point to the observer, to purely vertical at the furthest point ("on the other side of the black hole"). This different dependence of the shift vector with orbital phase depending on inclination has a considerable impact on the Schwarzschild QU loop asymmetry, see text for details.}
\label{fig:QUsym}
\end{figure*}

Let us consider a hot spot in Schwarzschild spacetime and the wavevector connecting this hot spot to the distant observer. The direction of this wavevector differs from the Minkowski case due to the existence of light bending. Let us write 
\be
\mathbf{k^S} \approx \mathbf{k^M} + \boldsymbol{\delta}\mathbf{k}^\mathrm{lensing}
\ee
where $\mathbf{k^S}$ is the Schwarzschild wavevector,
$\mathbf{k^M}$ is the Minkowski wavevector, and $\boldsymbol{\delta}\mathbf{k}^\mathrm{lensing}$ is the shift due to light bending. We note that this equation is not rigorous in the sense that we compare vectors that belong to tangent spaces to different manifolds, but it is still useful to get an intuition of the effect of light bending. The situation is illustrated in Fig.~\ref{fig:QUsym}, for face-on and edge-on inclinations. The lensing shift vector is a radial vector constant with orbital phase at zero inclination. This means that light bending does not break the QU loop mirror symmetry at zero inclination. Indeed, there are only three quantities that impact the Stokes parameters, namely
\begin{itemize}
\item the photon's energy in the emitter's frame, $\omega = -\mathbf{k} \cdot \mathbf{u}$, 
\item the cosine of the direction of emission in the emitter's frame, $\cos \theta_B = \mathbf{k} \cdot \mathbf{B}/\omega$, \footnote{It is clear from Eq.~\ref{eq:Kvec} that $\mathbf{K} \cdot \mathbf{B} = \mathbf{k} \cdot \mathbf{B}$.} 
\item the EVPA.
\end{itemize}
These quantities are independent of orbital phase at zero inclination, because of the constancy of the lensing shift vector with orbital phase, illustrated in the left panel of Fig.~\ref{fig:QUsym}. However, at edge-on inclination, the situation is completely changed and the lensing shift vector becomes very dependent on the orbital phase (see the right panel of Fig.~\ref{fig:QUsym}). This will lead to a strong dependence with orbital phase of the three quantities discussed above, and to the breaking of the QU loop mirror symmetry. This is in perfect agreement with the results of Fig.~\ref{fig:QUcompare} which shows that the loop mirror symmetry still holds at low inclination and becomes less and less conserved with increasing inclination. 
%We note that at very high inclination ($\iota \gtrsim 80^\circ$), the toroidal Schwarzschild QU loops would also become non-symmetric. 

\begin{figure}[htbp]  
\centering  
\includegraphics[width=0.5\textwidth]{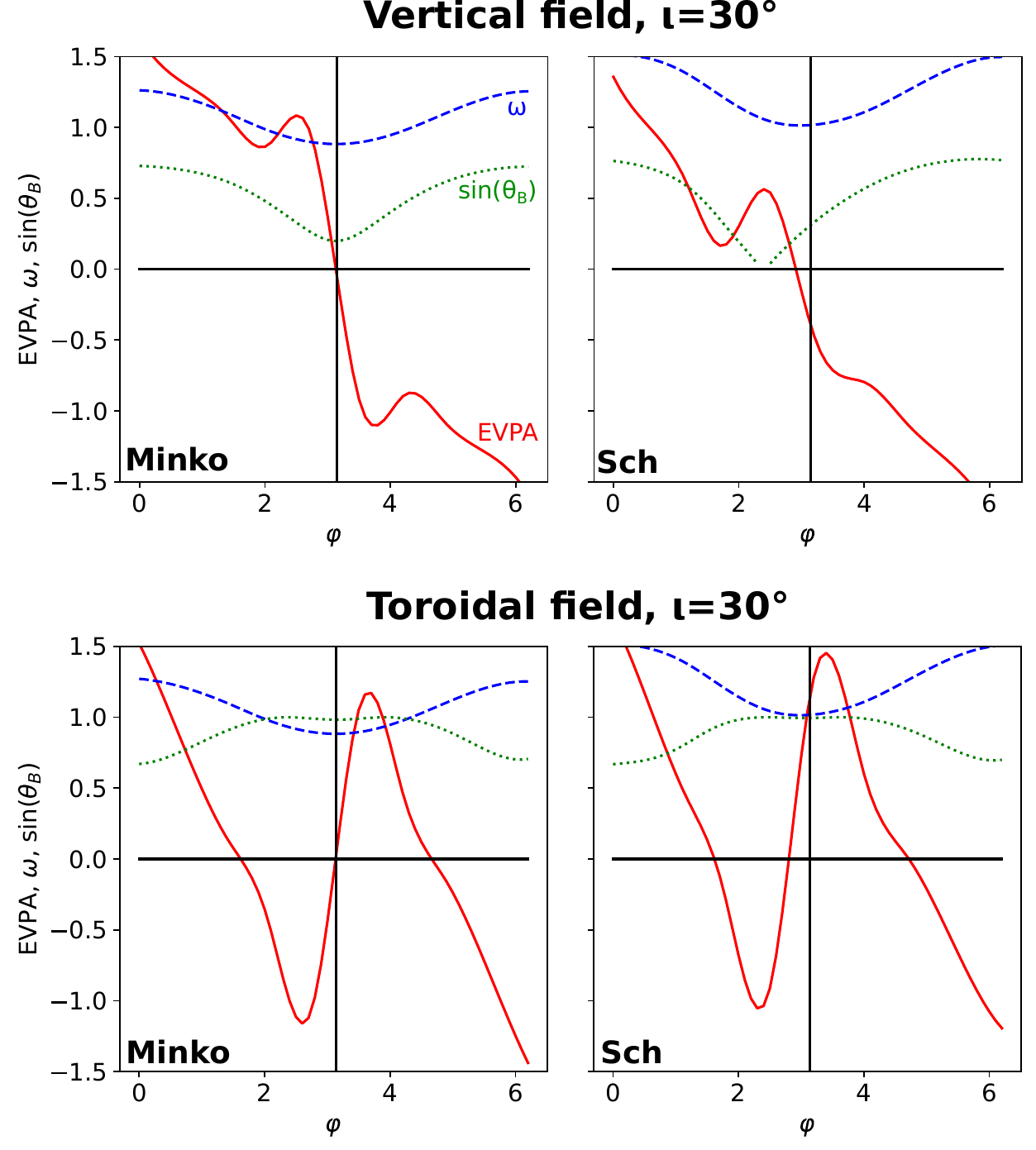}  
\caption{Evolution of orbit-varying quantities for Stokes parameters. 
The inclination is $\iota=30^\circ$.
The various panels show the evolution with orbital phase $\pp$ of the various quantities that impact the Stokes parameters: the EVPA (red), the photon's emitted energy $\omega = -\mathbf{k}\cdot \mathbf{u}$ (red, in units of the observed energy), and the sine of the emission angle $\sin \theta_B$ (green). The left column is computed in Minkowski, the right column in Schwarzschild. The top row is computed for a vertical magnetic field, the bottom row for a toroidal field. }
\label{fig:Qty_QUsym}
\end{figure}

One point remains to be discussed, which is why the QU loop mirror symmetry is broken much quicker with increasing inclination for a vertical magnetic field (in this case, the symmetry is lost already at $\iota \approx 30^\circ$) rather than for a toroidal field (in this case, the symmetry approximately holds until $\iota \approx 80^\circ$). This is related to the orbital phase evolution of the three quantities listed above. Figure~\ref{fig:Qty_QUsym} shows the orbital phase evolution of these quantities at $\iota=30^\circ$ in Minkowski and Schwarzschild, and for a vertical or toroidal field. This figure demonstrates that the EVPA and the emission direction are much more asymmetric for a vertical field than for a toroidal field, for this moderate inclination. 
In the toroidal case, the evolution of these quantities, although shifted in phase compared to the Minkowski case, remains rather similar to the flat-spacetime setup. We have checked that computing the QU track of a Schwarzschild/vertical setup at $\iota=30^\circ$, but imposing by hand some ad-hoc symmetric evolution of the EVPA and of the emission direction, leads to a mirror symmetric QU loop.

\section{Conclusion}
\label{sec:conclu}

This article has two main goals: (i) highlighting the role of special-relativistic aberration in generating the observed QU loops; (ii) elucidating an observable feature directly produced by spacetime curvature.

First, we highlight the crucial importance of special-relativistic effects in generating the observable QU loops associated with the polarized synchrotron flares of Sgr~A*. We have shown that most features discussed so far in the literature regarding QU loops (existence of the loops, number of loops, dependence with inclination and orbital radius) are already present in the Minkowski spacetime and are thus independent of light bending. The simplicity of Minkowski spacetime is a great asset allowing to develop a complete understanding of these features. 

Second, we indicate a specific property that is due to light bending. Minkowski QU loops are always mirror symmetric in the sense that the two half orbits lead to the same QU track. The axis of symmetry corresponds to the horizontal Q axis in our configuration with the angular momentum of the hot spot projected onto the observer's screen aligning with the vertical direction. In general the argument pertains to existence of any line of mirror symmetry in the QU plane, following the uncertain orientation of the observed system. %up to a symmetry with respect to the horizontal Q axis. 

On the contrary, and due to light bending, Schwarzschild QU loops are not symmetric in general. Schwarzschild QU loops in a toroidal magnetic field remain approximately (meaning to a better accuracy than current observations could tell) symmetric up to very high inclination (within $\approx 10^\circ$ of edge-on view). Nonetheless, Schwarzschild QU loops in a vertical magnetic field, which is the favored configuration for the likely MAD Sgr~A* flow, quickly lose their mirror symmetry with increasing inclination, and are already clearly asymmetric at a moderate inclination of about $30^\circ$. Thus, the asymmetry of the QU loops might constitute a compelling probe enabling quantification of the spacetime curvature in the close environment of Sgr~A*. {The detailed future studies of the QU loops could also constitute a path to confirming the existence of secondary images around black holes, which is another way to characterize curved spacetimes.}
%from analyzing current and future observed QU loops of Galactic center flares.

It is important to keep in mind the simplicity of our modeling and that astrophysical complexity might obscure the spacetime curvature effect on the asymmetry of the observed loop. A non-axisymmetric profile of the physical quantities (density, magnetic field, temperature) along the hot spot orbit might break the QU loop mirror symmetry even in the absence of curvature. {Internal physics of the hot spot (e.g., cooling) may have a similar effect by introducing time-dependence to the emission coefficient.} Non-circular motion, like an ejection along a jet sheath might also impact the conclusion. These possible limitations should be addressed in future works.
%\fv{Add caveat regarding the observability of the asymmetry + Faraday rotation in the millimeter.}

\appendix
\section{Analytical expressions in Minkowski spacetime}
\label{app:Minko}

Let us reiterate the expressions of the emitter's 4-velocity (Eq.~\ref{eq:uuM})
\be
\mathbf{u} = A \left(\boldsymbol{e_t} + r_0^{-1/2} \, \boldsymbol{e_\pp} \right),\quad A=\sqrt{\frac{r_0}{r_0-M}},
\ee
that of the wavevector (Eq.~\ref{eq:k4vec})
\be
  \mathbf{k} = \mathbf{e_t} - \sin \iota\, \sin \pp \,\mathbf{e_r} + \cos \iota \,\boldsymbol{e_\theta} - \sin \iota \, \cos \pp \,\boldsymbol{e_\pp}, 
\ee
that of the photon's emitted energy (Eq.~\ref{eq:om})
\be
\omega = -\mathbf{k} \cdot \mathbf{u} = A \left(1 + \frac{\sin \iota \cos \pp}{\sqrt{r_0}} \right),
\ee
that of the projection of $\mathbf{k}$ orthogonal to $\mathbf{u}$,
\begin{align}
  \mathbf{K} &= \mathbf{k} + \left( \mathbf{k} \cdot \mathbf{u} \right) \, \mathbf{u}
 \\ \nn
  &= \left( 1 - \omega A \right)\mathbf{e_t} - \sin \iota\, \sin \pp \,\mathbf{e_r} + \cos \iota \,\boldsymbol{e_\theta}  \\ \nn
  & \hspace{0.5cm}- \left( \sin \iota \, \cos \pp + \omega A r_0^{-1/2}  \right) \,\boldsymbol{e_\pp}, \\ \nn
\end{align}
and that of the observer's basis vectors (Eq.~\ref{eq:basis_vec})
\begin{align}
  \mathbf{e_w} &= \cos \pp \, \boldsymbol{e_r} - \sin \pp \,\boldsymbol{e_\pp}, \\ \nn
  \boldsymbol{e_\delta} &= -\cos \iota \sin \pp \, \mathbf{e_r} - \sin \iota \, \boldsymbol{e_\theta} - \cos \iota \cos \pp \, \boldsymbol{e_\pp}.  \\ \nn
\end{align}

\subsection{Vertical magnetic field}

Considering a unit vertical magnetic field
\be
\mathbf{\bar{B}} = -\boldsymbol{e_\theta},
\ee
we have
\be
\mathbf{\bar{B}} \cdot \mathbf{K} = - \cos \iota,
\ee
and
the projection of $\mathbf{\bar{B}}$ normal to the unit vector $\mathbf{\bar{K}} = \mathbf{K}/\omega$ along $\mathbf{K}$ reads
\begin{align}
\mathbf{B_\perp} &= \mathbf{\bar{B}} - \frac{\mathbf{\bar{B}} \cdot {\mathbf{{K}}}}{\omega^2} \, \mathbf{{K}} \\ \nn
&= \frac{\cos \iota}{\omega^2} \left[\left( 1 - \omega A \right)\mathbf{e_t} - \sin \iota\, \sin \pp \,\mathbf{e_r} + \left(\cos \iota - \frac{\omega^2}{\cos \iota}\right) \,\boldsymbol{e_\theta} \right. \\ \nn
& \hspace{1.5cm}\left. - \left( \sin \iota \, \cos \pp + \omega A r_0^{-1/2}  \right) \,\boldsymbol{e_\pp} \right].
\end{align}
The projections of this vector along the observer's basis vectors then read
\begin{align}
\mathbf{B_\perp} \cdot \mathbf{e_w} &=   \cos \iota \sin \pp \,\frac{A}{\omega \sqrt{r_0}}, \\ \nn
\mathbf{B_\perp} \cdot \boldsymbol{e_\delta}
&= \sin \iota + \cos^2 \iota \cos \pp \,\frac{A}{\omega \sqrt{r_0}}, \\ \nn
\end{align}
from which the EVPA expression of Eq.~\ref{eq:EVPAana} follows.

We also have
\be
\mathbf{\bar{K}} \cdot \mathbf{\bar{B}} = \frac{\mathbf{{K}}}{\omega} \cdot \mathbf{\bar{B}} =  - \frac{\cos \iota}{\omega}
\ee
where $\mathbf{\bar{K}}$ is the unit vector along $\mathbf{{K}}$.
We thus find the result of Eq.~\ref{eq:KB_vert}.

\subsection{Toroidal magnetic field}

Considering now a toroidal magnetic field
\be
\mathbf{\bar{B}} = A \left(\frac{\mathbf{e_t}}{\sqrt{r_0}} + \boldsymbol{e_\pp} \right),
\ee
we have
\be
\frac{\mathbf{\bar{B}} \cdot \mathbf{K}}{\omega^2} = - \frac{A}{\omega^2} \left( \frac{1}{\sqrt{r_0}} + \sin \iota \cos \pp\right) \equiv -C A
\ee
where we introduce
\be
C \equiv \frac{1}{\omega^2} \left(\frac{1}{\sqrt{r_0}} + \sin \iota \cos \pp \right).
\ee
So we get
\begin{align}
\mathbf{B_\perp} &= \mathbf{\bar{B}} - \frac{\mathbf{\bar{B}} \cdot {\mathbf{{K}}}}{\omega^2} \, \mathbf{{K}} \\ \nn
&= A \, \left[\frac{1}{\sqrt{r_0}} + C \left(1 - \omega A \right) \right] \,\mathbf{e_t} -  A C \, \sin \iota\, \sin \pp \,\mathbf{e_r} + AC \, \cos \iota \,\boldsymbol{e_\theta} \\ \nn
& \hspace{1.5cm} + A \, \left[ 1 - C \left( \sin \iota \, \cos \pp + \frac{\omega A}{ \sqrt{r_0}}  \right) \right]\,\boldsymbol{e_\pp}.
\end{align}
The projections onto the observer's basis vectors then read
\begin{align}
\mathbf{B_\perp} \cdot \mathbf{e_w} &=   A \sin \pp \left( C \, \frac{A \omega}{\sqrt{r_0}} - 1 \right), \\ \nn
\mathbf{B_\perp} \cdot \boldsymbol{e_\delta}
&= A \cos \pp \cos \iota \left( C \, \frac{A \omega}{\sqrt{r_0}} - 1 \right), \\ \nn
\end{align}
from which the EVPA expression of Eq.~\ref{eq:EVPA_tor} follows.
We also have
\be
\mathbf{\bar{K}} \cdot \mathbf{\bar{B}} = -C \omega A,
\ee
which is the result of Eq.~\ref{eq:KB_tor}.

%---------------------------------------------------------------------
%---------------------------------------------------------------------
\section*{Acknowledgements}
{
We thank Frank Eisenhauer, Jack Livingston, and Diogo Ribeiro for helpful comments to the draft.
This research is supported by the European Research Council advanced grant “M2FINDERS - Mapping Magnetic Fields with INterferometry Down to Event hoRizon Scales” (Grant No. 101018682).
  
  }
%---------------------------------------------------------------------
%-------------------------------
 % --------------------------------------

\bibliography{QUloop}
\bibliographystyle{aa}

\end{document}